\documentclass[aps,prb,superscriptaddress,twocolumn,showpacs]{revtex4}

\usepackage{amsmath,amssymb}
\usepackage{graphicx}
\usepackage{feynmf}
\usepackage{array}
\usepackage{color}	
\usepackage{verbatim}	

\newcommand{\rs}{\rm \scriptscriptstyle}

\newcommand{\rsF}{{\rm \scriptscriptstyle F}}
\newcommand{\M}{\mathcal{M}}
\newcommand{\A}{\mathcal{A}}
\newcommand{\B}{\mathcal{B}}
\newcommand{\D}{\mathcal{D}}

\begin{document}

\title{Projective versus weak measurement of charge in a mesoscopic
conductor}

\author{D. Oehri}
\affiliation{Theoretische Physik,
  ETH Zurich, CH-8093 Zurich, Switzerland}

\author{A.V. Lebedev}
\affiliation{Theoretische Physik,
  ETH Zurich, CH-8093 Zurich, Switzerland}

\author{G.B. Lesovik}
\affiliation{L.D.\ Landau Institute for Theoretical Physics RAS,
   117940 Moscow, Russia}

\author{G. Blatter}
\affiliation{Theoretische Physik,
  ETH Zurich, CH-8093 Zurich, Switzerland}

\date{\today}

\begin{abstract}
We study the charge dynamics of a quantum dot as measured by a nearby quantum
point contact probing the dot via individual single-particle wave packets.  We
contrast the two limiting cases of weak and strong system--detector coupling
exerting vanishing and strong back-action on the system and analyze the
resulting differences in the charge-charge correlator.  Extending the study to
multiple projective measurements modelling a continuous strong measurement, we
identify a transition from a charge dynamics dominated by the system's
properties to a universal dynamics governed by the measurement.
\end{abstract}


\maketitle

%%%%%%%%%%%%%%%%%%%%%%%%%%%%%%%%%%%%%%%%%%%%%%
\section{Introduction}
%%%%%%%%%%%%%%%%%%%%%%%%%%%%%%%%%%%%%%%%%%%%%%

Discussions on quantum measurement \cite{braginsky:92,wiseman:10,clerk:10}
usually rely on two fundamental elements, the Born rule \cite{born:26},
telling us how to extract information from the quantum mechanical wave
function, and the von Neumann \cite{neumann:31} projection postulate, stating
that measuring a system observable generates a collapse of the wave function
and telling us how to restart the system's unitary evolution after the
measurement. A third most important element is that of the detector's
back-action on the system which is at the heart of the projection process when
viewed from a microscopic perspective; much effort has gone into the
understanding of the phenomenological von Neumann projection in terms of a
unitary evolution of the entangled system--detector dynamics. Here, we take a
step back and study the charge dynamics, specifically the charge-charge
correlator, of a quantum dot (QD) as measured by a quantum point contact (QPC)
in order to understand the impact of the detector's back-action on the time
evolution of the correlator. We study the impact of the back-action in two
limiting cases: i) a weak detector--system coupling which we treat
perturbatively describing the limit of no back-action and ii) an
intermediate/strong system--detector coupling which we describe by a von
Neumann projection accounting for the limit of strong back-action. We
determine the physically measureable correlators and quantitatively analyze
their difference for the case of a quantum dot with a single resonant level.

Understanding quantum measurement as the bridge between the quantum- and our
classical work is a fascinating and broad topic, ranging from fundamental
aspects of the measurement problem \cite{ever:57,griff:01,whee:83,zurek:03} to
such practical issues as optimizing the information gain at minimal system
invasion \cite{ave:05}.  Much effort has gone into the microscopic
understanding of the measurement process and its interrelation with
back-action, treating quantum averaged evolutions
\cite{alei:97,levi:97,gur:97,stodo:99,shni:98,makhlin:00}, selective system
dynamics \cite{korot:99,korot:01}, and correlated weak-strong measurements in
the form of weak values \cite{ahar:88,dressel:14,romito:08,zilber:14}.
Quantum dots in transport \cite{alei:97,levi:97,zilber:14,buks:98,gust:06},
isolated double quantum dots
\cite{gur:97,stodo:99,korot:99,korot:01,romito:08}, as well as quantum point
contacts \cite{hasko:93,vander:04,pilgram:02,clerk:03} have played a central
role in analyzing quantum measurement within the realm of mesoscopic physics.

In the present study, we focus on the charge dynamics of a quantum dot (or a
localized region of a mesoscopic scatterer in more general terms) as measured
by a nearby QPC detector and determine the time evolution of the average
charge and the charge-charge correlator for different strengths of the
system--detector coupling. Thereby we keep in mind a measurement with
single-particle wave packets\cite{feve:07,dubois:13} incident on the detector
at times $t_1$ and $t_2$ with their reflection/transmission through the QPC
providing information on the charge state of the quantum dot at the two time
instances.  Our analysis provides us with two central results: on a technical
level, we find that the strong projective measurement can be expressed through
a projected charge $\hat{Q}^P(t|t_0)$, the charge at time $t$ after a previous
projection at time $t_0$, for which we find a compact expression in terms of
the system's scattering matrix. On a physical level, we find that, in spite of
the strongly different back-action induced by the first measurement, the two
correlators for weak and strong measurements come out qualitatively similar,
although quantitative differences remain, of course; the latter are
specifically discussed for the single-resonance level model.  Furthermore, we
attempt to model the case of a finite constant voltage $V$ applied to the QPC
detector by considering a sequence of wave packets incident on the QPC,
leading to repeated projections which we describe via the projection
postulate. Increasing the rate of projections, we identify a transition from a
regime where the dynamics of the system is dominated by the system's
characteristics to a regime where the dynamics is universal and uniquely
determined by the measurement.

The paper is organized as follows: In Sec.\ \ref{sec:setup} we introduce the
model describing our mesoscopic conductor and discuss the different regimes of
projective (intermediate/strong coupling) and weak measurements, identifying
the measurable charge-charge correlator in each of these regimes. In Sec.\
\ref{sec:cc}, the two charge-charge correlators are calculated for arbitrary
scatterers and their difference is discussed quantitatively for the case of a
single-level quantum dot. In Sec.\ \ref{sec:mp}, the discussion is extended to
the case of repeated projective measurements. In order to understand the
universal behavior found for fast repeated measurements, we consider the model
of a fluctuating quantum dot level. A summary and conclusions are given in
Sec.\ \ref{sec:con}.

%%%%%%%%%%%%%%%%%%%%%%%%%%%%%%%%%%%%%%%%%%%%%%
\section{Formalism}\label{sec:setup}
%%%%%%%%%%%%%%%%%%%%%%%%%%%%%%%%%%%%%%%%%%%%%%

\begin{figure}[h]
\begin{center}
\includegraphics[width=6cm]{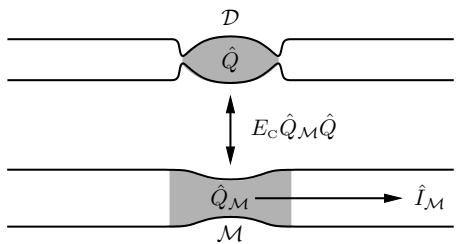}
\caption{The system of interest with the central region $\mathcal{D}$ holding
the charge $\hat{Q}$ (here a quantum dot) capacitively coupled to the central
region $\mathcal{M}$ of the measurement system. The current
$\hat{I}_\mathcal{M}$ in the detector serves as a readout for the charge
$\hat{Q}$ in the region $\mathcal{D}$.}
\label{fig:sysdet}
\end{center}
\end{figure}

We consider a mesoscopic conductor with a central scattering region (e.g., a
quantum dot), described through its single-particle scattering matrix ${\bf
S}_{k}$ (see Ref.\ \onlinecite{lesovik:11} for a review on the scattering
matrix approach to mesoscopic transport). Here, we are interested in the
dynamics of the charge $Q(t)$ (measured in units of electronic charge $e$) in
the scattering region and its modifications when it is subjected to a (strong)
measurement.  In order to measure this dynamics and specifically the
associated charge-charge correlator, the scattering region is capacitively
coupled to the quantum point contact (QPC) of a detector system,
see Fig.\ \ref{fig:sysdet}. The charge dynamics $Q(t)$ of the system and the
current $I_\mathcal{M}(t)$ through the detector then mutually influence one
another and we have to analyze the measurement process in order to identify
the measurable charge-charge correlator. 
%

%%%%%%%%%%%%%%%%%%%%%%%%%%%%%%%%%%%%%%%%%%%%%%
\subsection{The model system}\label{sec:mod}
%%%%%%%%%%%%%%%%%%%%%%%%%%%%%%%%%%%%%%%%%%%%%%

We consider a one-dimensional non-interacting system consisting of two
half-infinite leads connected through a central region $\mathcal{D} =
[-d/2,d/2]$ where particles are scattered by the single-particle potential
$\hat{V}$. The latter is characterized by the single-particle scattering
matrix
\begin{align}
{\bf S}_k=
\begin{pmatrix}
r_{L k} 	& 	t_{k} \\
t_k 		&	r_{Rk}
\end{pmatrix}.
\end{align}
We make use of the Lippmann-Schwinger scattering states $|\varphi_{a
k}\rangle$ satisfying $(\hat{H}_{\rm kin}+\hat{V})\, |\varphi_{ak}\rangle =
\epsilon_k\, |\varphi_{ak}\rangle$ with the kinetic part of the Hamiltonian
$\hat{H}_{\rm kin}$, the wavevector $k>0$ and $a=L/R$ describing a scattering
state incoming from the left/right.  We linearize the spectrum around the
Fermi energy $\epsilon_{\rsF}$, i.e., $\epsilon_k = \epsilon_{\rsF} + \hbar
v_{\rsF} (k-k_{\rsF})$. The asymptotics $|x| \to \infty$ of the (LS)
scattering states is described by the scattering amplitudes, i.e.,
\begin{align}
   \varphi_{Lk}
    &\sim \Theta(-x)(e^{ikx}+r_{Lk}e^{-ikx})+\Theta(x)t_{k}e^{ikx},
\nonumber\\
   \varphi_{Rk}
    &\sim \Theta(-x)t_{k}e^{-ikx}+\Theta(x)(e^{-ikx}+r_{Rk}e^{ikx}).
\label{eq:asympLS}
\end{align}
Spin is trivially accounted for in our non-interacting system and hence we
restrict ourselves to spinless particles.

As shown in Refs.\ \onlinecite{clerk:03} and \onlinecite{oehri:12}, the
operator $\hat{Q}$ describing the charge in the central region $\mathcal{D}$
(in units of the electronic charge $e$) can be expressed through creation
(annihilation) operators $\hat{c}^\dagger_{ak}$ ($\hat{c}^{\phantom
\dagger}_{ak}$) of the above Lippmann-Schwinger scattering states via
\begin{align}
   \hat{Q}(t)&= \sum_{a^\prime a}
   \int\frac{dk^\prime}{2\pi}\int\frac{dk}{2\pi}
   A_{a^\prime k^\prime, ak}(t)
   \hat{c}^\dagger_{a^\prime k^\prime} \hat{c}^{\phantom\dagger}_{ak}
   \label{eq:Qexpr}\\
   &= \sum_{\alpha'\alpha} A_{\alpha',\alpha}(t)
   \hat{c}^\dagger_{\alpha'} \hat{c}^{\phantom{\dagger}}_{\alpha},
\label{eq:Qsexpr}
\end{align}
with the notation $\alpha=(a,k)$ and $\sum_{\alpha} = \sum_a \int(dk/2\pi)$.
The matrix elements $A_{\alpha',\alpha}(t)= [{\bf A}_{k^\prime,k}
(t)]_{a^\prime, a}$ are related to the scattering matrix $\tilde{\bf S}_k={\bf
S}_k e^{ikd}$,
\begin{align}
   {\bf A}_{k^\prime,k}(t)
   = -i\frac{{\bf 1}-\tilde{{\bf S}}^\dagger_{k^\prime}
     \tilde{{\bf S}}^{\phantom{\dagger}}_{k}}{k^\prime-k}
     e^{i(k^\prime-k)(v_{\rsF}t+d/2)}
\label{eq:A}
\end{align}
with ${\bf A}_{k,k}(t) = -i\tilde{{\bf S}}^\dagger_{k} \partial_k \tilde{{\bf
S}}^{\phantom{\dagger}}_{k}$ and where we have dropped terms of order
$\mathcal{O}(1/k_\rsF)$. The matrix ${\bf A}$ is equivalent to the density of
states matrix expression introduced in Ref.\ \onlinecite{pedersen:98} and was
used in Ref.\ \onlinecite{oehri:12} to express the interaction kernel through
the scattering states in a discussion of interacting electron transport; 
here, it is used to express the charge
through the scattering states, see Eq.\ \eqref{eq:Qsexpr}, and to obtain
simple and compact expressions for the projected charge, see Eqs.\
\eqref{eq:QP} and \eqref{eq:AP}. With regard to the latter, we note that the
matrix ${\bf A}$ possesses the useful projector property
\begin{align}
\int\frac{dp}{2\pi}{\bf A}_{k,p}(t){\bf A}_{p,q}(t)={\bf A}_{k,q}(t),
\label{eq:AAA}
\end{align}
a consequence of the projector property $\hat{Q}(t) \hat{Q}(t) = \hat{Q}(t)$
of the charge operator restricted to the single-particle Hilbert space; the
latter is easily checked by acting (twice) with the charge operator
$\hat{Q}(t) \equiv \hat{\mathcal{P}}_{\D}(t) = \int_{\mathcal{D}}dx\,
\hat{\psi}^\dagger (x,t) \hat{\psi}(x,t)$ on a single-particle state
$|\Psi_1\rangle$ (here, $\hat{\mathcal{P}}_{\D}(t)$ denotes the real-space
projector on the region $\D$).  The transport through $\mathcal{D}$ is
implemented by connecting the two semi-infinite leads to two reservoirs at
chemical potential $\mu_{L}$ and $\mu_{R}$ with $\mu_{L/R}=\epsilon_{\rsF}\pm
eV/2$ as described by the steady state density matrix $\hat{\rho}_0$.

Within this model, we are able to describe local properties of
the system, specifically the charge dynamics in the region $\mathcal{D}$,
through the single-particle scattering matrix ${\bf S}_k$ which is an
asymptotic property of the system. Note that interactions are limited to the
system--detector coupling, i.e., we are considering a non-interacting system;
interactions within the system could be taken into account within perturbation
theory \cite{oehri:12}.

%%%%%%%%%%%%%%%%%%%%%%%%%%%%%%%%%%%%%%%%%%%%%%
\subsection{The detector}\label{sec:regimes}
%%%%%%%%%%%%%%%%%%%%%%%%%%%%%%%%%%%%%%%%%%%%%%

\begin{figure}[ht]
\begin{center}
\includegraphics[scale=.95]{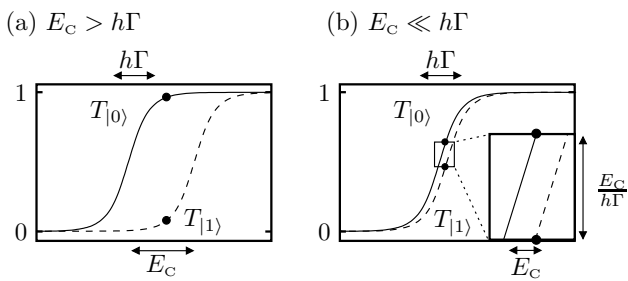}
\caption{(a) The QPC transmission changes from 0 to 1 within a width
$h\Gamma$. In the case of strong coupling $E_{\rs C}>h\Gamma$, the change of
transmission between $T_{|0\rangle}$ (empty dot) and $T_{|1\rangle}$ (filled
dot) can be tuned to about unity (see black dots). (b) For weak coupling,
$E_{\rs C}\ll h\Gamma$, the change of transmission (see black dots) is small,
i.e., $\Delta T\approx E_{\rs C}/h\Gamma\ll 1$.}
\label{fig:meas_reg}
\end{center}
\end{figure}

In order to measure the system charge $\hat{Q}$ we make use of a capacitively
coupled QPC detector \cite{hasko:93}. The latter is characterized by a
step-like transmission characteristic, see Fig.\ \ref{fig:meas_reg}, with a
width $h\Gamma$ related to the tunneling time $t_\mathrm{tun} \sim 1/\Gamma$
of particles traversing the constriction. The coupling $\hat{H}_{\rm
coupl}=E_{\rs C} \hat{Q}_{\mathcal{M}} \hat{Q}$ between the system charge
$\hat{Q}$ and the charge $\hat{Q}_{\mathcal{M}}$ in the QPC region
$\mathcal{M}$ of the detector will shift the location of the transmission step
and the detector current will provide information about the system, while at
the same time cause an unavoidable back-action.

To fix ideas, we restrict the discussion to a single-level quantum dot (QD)
with two charge states $|0\rangle$ and $|1\rangle$ and focus on the limit
where each detector electron probes the dot individually, requiring that the
tunneling time $t_{\rm tun} \sim 1/\Gamma$ is small compared to the typical
separation between detector electrons (alternatively, we can consider a weakly
driven detector with bias voltage $eV \ll h\Gamma$ such that the typical time
separation $t_{V} \sim h/eV > t_{\rm tun}$ between electrons is large).  In
order for the detector to measure the system it should be fast, $t_{\rm tun}
\ll t_{\rm sys}$, where $t_{\rm sys}$ is a typical system timescale.  In this
situation, the system is in a fixed charge state during the tunneling of a
detector electron. If the QD is empty (state $|0\rangle$), the electron is
transmitted with probability $T_{|0\rangle}\approx T(\epsilon_{\rs F})$, while
for a filled QD (state $|1\rangle$), the transmission probability is
$T_{|1\rangle} \approx T(\epsilon_{\rs F}-E_{\rs C})$. We then identify two
measurement regimes: i) a strong coupling regime with $E_{\rs C} > h\Gamma$,
see Fig.\ \ref{fig:meas_reg}(a), where we can choose a working point such that
$T_{|0\rangle}\approx 1$ and $T_{|1\rangle}\approx 0$, and ii) a weak coupling
regime $E_{\rs C}< h\Gamma$, see Fig.\ \ref{fig:meas_reg}(b), where the change
in transmission $\Delta T=T_{|0\rangle}-T_{|1\rangle}\approx E_{\rs
C}/h\Gamma\ll1$ is small.

%%%%%%%%%%%%%%%%%%%%%%%%%%%%%%%%%%%%%%%%%%%%%%%%%%%%%%%%%%%%%%%%%%%%%%%
\subsection{Projective measurement}\label{sec:proj}
%%%%%%%%%%%%%%%%%%%%%%%%%%%%%%%%%%%%%%%%%%%%%%%%%%%%%%%%%%%%%%%%%%%%%%%

For strong coupling, $E_{\rs C} > h\Gamma$ the charge states of the dot and
the state of one single QPC electron after scattering [i.e., transmitted
($|t\rangle$) or reflected ($|r\rangle$)] become fully entangled, $\alpha
|0\rangle \otimes |t\rangle + \beta |1\rangle \otimes |r\rangle$. The
detection of this electron leads to the `collapse' of the dot state and fully
determines the charge, such that the measurement with even a single QPC
electron is a strong one, able to acquire the full information; the
concomitant projection of the dot state corresponds to the strongest possible
back-action of the detector on the system.  Hence, in order to measure the
charge $\hat{Q}(t)$ at time $t$ at strong coupling $E_{\rs C} > h\Gamma$, one
single electron \cite{feve:07,dubois:13} is sent towards the QPC to arrive there at time
$t$.  The detection of the transmitted electron determines the charge state of
region $\D$ and projects the system to the corresponding state.  Repeating
this measurement (after the system has equilibrated) allows to determine the
average charge $Q$.  Similarly, in order to measure the charge-charge
correlator, the dot charge has to be measured at times $t_1$ and $t_2$ and the
measurement has to be repeated after equilibration at fixed time delay $\Delta
t=t_2-t_1$ in order to find the average charge-charge correlator for times
$t_1$ and $t_2$.

In order to describe this correlator theoretically, we describe the
measurement using the projection postulate \cite{neumann:31}: the measurement
of the charge operator $\hat{Q}$  at $t$ projects the system onto a state with
an integer number of charges in the central region $\mathcal{D}$ which we
account for by the operator
\begin{align}
   \hat{P}_N(t)=\int_0^{2\pi}\frac{d\lambda}{2\pi} e^{i\lambda(\hat{Q}(t)-N)}
\label{eq:PN}
\end{align}
projecting the state of the system at time $t$ onto a state with $N$ charges
in the central region $\mathcal{D}$.  While the system's state for times
$t'<t$ is described by the steady-state density matrix $\hat{\rho}_0$, the
charge measurement at time $t$ projects the density matrix onto well-defined
charge states which are expressed through the operator $\hat{P}_N$ as $\sum_N
\hat{P}_N(t) \hat{\rho}_0 \hat{P}_N(t)$. Then the {\it projected}
charge-charge correlator at times $t_1$ and $t_2>t_1$ is given by
\begin{align}
   S^{P}_{QQ}(t_2,t_1)
   =\sum_N \text{Tr}\Bigl\{
   \hat{Q}(t_2)\hat{Q}(t_1)
   \hat{P}_{N}(t_1)\hat{\rho}_0\hat{P}_{N}(t_1)
   \Bigr\},
\end{align}
where the collapse of the state at time $t_1$ is accounted for by the
projection operators $\hat{P}_N(t_1)$ while the measurement at time $t_2$ does
not affect the result. Making use of the cyclic property of the trace and
$[\hat{Q}(t_1), \hat{P}_N(t_1)] = 0$, we obtain the projected correlator
\begin{align}\label{eq:SPQQ}
   S^{P}_{QQ}(t_2,t_1)
   = \langle \hat{Q}^{P}(t_2|t_1)\hat{Q}(t_1) \rangle,
\end{align}
with $\langle\hat{O}\rangle=\text{Tr}\{\hat{O}\hat{\rho}_0\}$ and the operator
$\hat{Q}^{P}(t_2|t_1)$ describing the charge in the central region at time
$t_2$ provided that the state of the system was projected at time $t_1<t_2$,
\begin{align}
   \hat{Q}^{P}(t_2|t_1)=\sum_N \hat{P}_{N}(t_1)\hat{Q}(t_2)\hat{P}_{N}(t_1).
\label{eq:QPdef}
\end{align}
Eq.\ \eqref{eq:SPQQ} corresponds to the measurable charge-charge correlator in
the regime of strong, projective measurements. Taking into account the spin
degree of freedom, Eq.\ \eqref{eq:SPQQ} remains unchanged with
$\hat{Q}(t_1)=\hat{Q}_{\downarrow}(t_1)+\hat{Q}_{\uparrow}(t_1)$ and
$\hat{Q}^P(t_2|t_1) = \hat{Q}_{\downarrow}^P(t_2|t_1) + \hat{Q}_{\uparrow}^P
(t_2|t_1)$; furthermore, the up- and down spin components of charge are
uncorrelated.

At intermediate coupling $E_{\rs C}<h\Gamma$ the scattering of one single
electron becomes a probabilistic process (with finite transmission and
reflection probability for both charge states) such that one electron alone
does not provide the information on the dot's state and many electrons are
required to probe the dot. The full information about the dot state is
acquired after the passage of $N$ electrons when the measured transmitted
charge $Q_{\rm tr}(N)$ through the QPC can be attributed to a particular dot
state, i.e., the difference $Q_{\rm tr}^{|0\rangle} (N) - Q_{\rm
tr}^{|1\rangle} (N) = \Delta T\, N$ has to be larger than the standard
deviation $\sigma$ of the probabilistic process of charge transmission. At
small temperatures, the latter is determined by the charge partitioning noise,
i.e., $\sigma^2 = T(1-T)\, N$, such that the required number of probe
electrons is \cite{alei:97,levi:97,gur:97,stodo:99}
\begin{align}
   N \sim \Bigl(\frac{h\Gamma}{E_{\rs C}}\Bigr)^2T(1-T).
   \label{eq:tms}
\end{align}
For the case where a finite voltage $V$ is applied across the QPC detector, we
obtain the measurement time $t_{\rm ms} \sim N t_V$ with $t_V = h/eV$ the
voltage time. For small temperatures $k_{\rs B} T < eV$
the partitioning noise dominates over the thermal noise.

Next, we have to account for the time scale $t_{\rm sys}$ of the dot. For
$t_{\rm ms} \ll t_{\rm sys}$, the state of the dot can be determined by
probing the dot for some measurement time larger than $t_{\rm ms}$; such a
measurement is strong and projective.  On the other hand, the measurement is
generically weak if $t_{\rm ms} > t_{\rm sys}$ and we have to find an
alternative procedure to find the information on the dot's charge state; we
will discuss this weak measurement regime in the next section.

The discussion in this section has been limited to the case of a single-level
quantum dot, however, the extension to the case of a more complex system with
more charge states is straightforward. While in such a situation it is not
possible to resolve all charge states by one electron alone, a projective
measurement involving many electrons or a weak measurement as described below
are still possible.

%%%%%%%%%%%%%%%%%%%%%%%%%%%%%%%%%%%%%%%%%%%%%%
\subsection{Weak coupling, weak measurement regime}\label{sec:weakc}
%%%%%%%%%%%%%%%%%%%%%%%%%%%%%%%%%%%%%%%%%%%%%%

At generically weak coupling $E_{\rs C} \ll h\Gamma$ (such that $t_{\rm ms} >
t_{\rm sys}$), the system cannot be measured during the system time
$t_\mathrm{sys}$ and we need an alternative measurement scheme. The capacitive
coupling between the system and the detector still affects the detector
current $\hat{I}_{\mathcal{M}}$ and thus can be used to learn about the
system's charge dynamics ${Q}(t)$.  The detector current is given by the
expectation value $\langle \hat{I}^H_{\mathcal{M}}(x,t)\rangle = {\rm
Tr}\{\hat{I}^H_{\mathcal{M}}(x,t) \, \hat{\rho}_\mathrm{sd}\}$, where
$\hat{I}^H_{\mathcal{M}}(x,t)$ is the current operator in the Heisenberg
representation and $\hat{\rho}_\mathrm{sd}$ is the steady-state density matrix
of the combined system--detector setup, $\hat{\rho}_\mathrm{sd} = \hat{\rho}_0
\otimes \hat{\rho}_0^{\M}$. For a weak system--detector coupling, we can
calculate the modulation in the detector current perturbatively
\cite{korot:01b,ave:03,clerk:03,jordan:05}; to lowest order in the coupling
strength $E_{\rs C}$ the result is
\begin{align}
\label{eq:Iexp}
   \langle \hat{I}^H_{\mathcal{M}}(x,t)\rangle
   &\approx \langle \hat{I}_{\mathcal{M}}(x,t)\rangle
   +\frac{2E_{\rs C}}{\hbar}\!\!\int\limits_{-\infty}^{t}\!\!
   d\tau\, \mathcal{I}S^{\mathcal{M}}_{IQ}(x,t;\tau)
   \langle\hat{Q}(\tau)\rangle,
\end{align}
where $\hat{I}_{\mathcal{M}}(x,t)$ is the current operator in the interaction
representation (i.e., with respect to the coupling Hamiltonian) and $\langle
\hat{\mathcal{O}}\rangle = {\rm Tr}\{\hat{\mathcal{O}} \,
\hat{\rho}_\mathrm{sd}\}$. In the expression above, $\mathcal{I}
S^{\mathcal{M}}_{IQ} (x,t;\tau)$ is the imaginary part of the correlator
between the detector current $\hat{I}_\mathcal{M}(x,t)$ and the detector
charge $\hat{Q}_\mathcal{M}(\tau)$, i.e., $\mathcal{I} S^{\mathcal{M}}_{IQ}
(x,t;\tau) = -i\langle\langle [\hat{I}_{\mathcal{M}} (x,t),
\hat{Q}_{\mathcal{M}} (\tau)] \rangle\rangle/2$ with $\langle
\langle\hat{\mathcal{O}}_1 \hat{\mathcal{O}}_2\rangle\rangle =
\langle\hat{\mathcal{O}}_1 \hat{\mathcal{O}}_2\rangle - \langle
\hat{\mathcal{O}}_1\rangle\langle \hat{\mathcal{O}}_2\rangle$. The quantity
$\mathcal{I}S^{\mathcal{M}}_{IQ} (x,t;\tau)$ is a response function of the
measurement apparatus $\mathcal{M}$ and hence a pure detector property,
independent of the system. A good detector is much faster than the dynamics of
the system of interest, i.e., $t_{\rm tun} \ll t_{\rm sys}$, such that the
charge $\langle\hat{Q}(\tau)\rangle$ can be treated as slowly varying on the
time scale of the detector. The response function
$\mathcal{I}S^{\mathcal{M}}_{IQ}$ then is effectively given by the zero
frequency response function, i.e., $\mathcal{I}
S^{\M}_{IQ}(x,t;\tau)\approx\mathcal{I}S^{\M}_{IQ,\omega=0}
\delta[\tau-(t-x/v_{\rsF})]$, such that
\begin{align}
\label{eq:Iexp2}
   \langle \hat{I}^H_{\mathcal{M}}(x,t)\rangle
   &\approx \langle \hat{I}_{\mathcal{M}}(x)\rangle
   +\frac{2E_{\rs C}}{\hbar}\mathcal{I}S^{\mathcal{M}}_{IQ,\omega=0}
   \langle\hat{Q}(t-x/v_{\rsF})\rangle.
\end{align}
This relation allows one to determine the time averaged charge expectation
value $\bar{Q}$ from an experimentally measured time trace of $I_{\M}(x,t)$ by
calculating the time averaged current $\bar{I}_{\M}=(1/T)\int_0^{T} dt \,
I_{\M}(x,t)$. 

Next, we derive an expression for the charge correlator at times $t_1$ and
$t_2$ in the weak-coupling regime.  We consider two detectors\cite{jordan:05}
$\mathcal{A}$ and $\mathcal{B}$ capacitively coupled to the system, i.e.,
$\hat{H}_{\rm coupl}=E_{{\rs C}}^\A \hat{Q}_{\A} \hat{Q}+E_{{\rs C}}^\B
\hat{Q}_{\B}\hat{Q}$. The correlation between the two detectors arises due to
the coupling between the two detectors via the system charge. The measurable
time-domain correlator of the detector is the {\it irreducible} symmetrized
current-current correlator\cite{lesovik:98} $\langle\langle \hat{I}^H_{\A}
(x_{\A},t_{\A})\hat{I}^H_{\B}(x_{\B},t_{\B}) + \hat{I}^H_{\B}
(x_{\B},t_{\B})\hat{I}^H_{\A} (x_{\A},t_{\A})\rangle\rangle/2$ (in contrast to
a measurement in the frequency domain, where a non-symmetrized correlator is
measured at positive frequencies \cite{lesovik:97,gavish:00,aguado:00}). The
above symmetrized detector correlator is related to the charge-charge
correlator of the dot via
\begin{widetext}
\begin{align}
   &\mathcal{R}S^{\rm irr}_{I_{\A}I_{\B}}(x_{\A},t_{\A};x_{\B},t_{\B})
   = \langle\langle\hat{I}^H_{\A}(x_{\A},t_{\A})\hat{I}^H_{\B}
   (x_{\B},t_{\B})+\hat{I}^H_{\B}(x_{\B},t_{\B})\hat{I}^H_{\A}
   (x_{\A},t_{\A})\rangle\rangle/2\nonumber\\
   &\hspace{20pt}\approx
   \frac{E_{{\rs C}}^\A E_{{\rs C}}^\B}{\hbar^2}
   \iint_{-\infty}^{\infty} d\tau_{\A}d\tau_{\B}
   \Bigl(\Theta(t_{\A}-\tau_{\A})\Theta(t_{\B}-\tau_{\B})
   \mathcal{I}S^{\A}_{IQ}(x_{\A},t_{\A};\tau_{\A})
   \mathcal{I}S^{\B}_{IQ}(x_{\B},t_{\B};\tau_{\B})
   \mathcal{R}S^{\rm irr}_{QQ}(\tau_A,\tau_B)\nonumber\\
   &\hspace{130pt}
   +\Theta(t_{\A}-\tau_{\A})\Theta(\tau_{\A}-\tau_{\B})
   \mathcal{I}S^{\A}_{IQ}(x_{\A},t_{\A};\tau_{\A})
   \mathcal{R}S^{\B}_{IQ}(x_{\B},t_{\B};\tau_{\B})
   \mathcal{I}S^{\rm irr}_{QQ}(\tau_{\A},\tau_{\B})\nonumber\\
   &\hspace{130pt}
   -\Theta(t_{\B}-\tau_{\B})\Theta(\tau_{\B}-\tau_{\A})
   \mathcal{R}S^{\A}_{IQ}(x_{\A},t_{\A};\tau_{\A})
   \mathcal{I}S^{\B}_{IQ}(x_{\B},t_{\B};\tau_{\B})
   \mathcal{I}S^{\rm irr}_{QQ}(\tau_{\A},\tau_{\B})
   \Bigr),
\end{align}
\end{widetext}
as follows from a perturbative analysis in the lowest order of the couplings.
In the above expression,
\begin{align}
   \mathcal{R}S^{\M}_{IQ}(x,t;\tau)&
   =\langle\langle\{\hat{I}_{\M}(x,t),\hat{Q}_{\M}(\tau)\}\rangle\rangle/2\\
   \mathcal{I}S^{\M}_{IQ}(x,t;\tau)&
   =-i\langle\langle[\hat{I}_{\M}(x,t),\hat{Q}_{\M}(\tau)]\rangle\rangle/2
\end{align}
are the real and imaginary parts of the current-charge correlator in the
detectors $\M=\A$ or $\B$ representing pure detector response functions.  On
the other hand, $\mathcal{R}S^{\rm irr}_{QQ} (\tau_{\A}, \tau_{\B}) =
\langle\langle \{\hat{Q} (\tau_{\A}), \hat{Q} (\tau_{\B})\} \rangle\rangle/2$
and $\mathcal{I} S^{\rm irr}_{QQ} (\tau_{\A}, \tau_{\B}) = -i\langle\langle
[\hat{Q}(\tau_{\A}), \hat{Q}(\tau_{\B})] \rangle\rangle/2$ are the real
(symmetrized) and imaginary (anti-symmetrized) part of the {\it irreducible}
charge-charge correlator of the scattering region $\D$ which we are interested
in. Again, for good, i.e., fast $t_{\rm tun} \ll t_{\rm sys}$, detectors, the
system quantities $\mathcal{I}S^{\rm irr}_{QQ}$ and $\mathcal{R}S^{\rm
irr}_{QQ}$ are slowly varying on the time scales of the detectors and the
response functions effectively are given by the zero frequency response
functions $\mathcal{I} S^{\M}_{IQ, \omega=0}$ and $\mathcal{R} S^{\M}_{IQ,
\omega=0}$. For the QPC detector with a symmetric scattering potential, the
real part of the zero frequency response vanishes \cite{pilgram:02,clerk:03},
i.e., $\mathcal{R} S^{\M}_{IQ, \omega=0} = 0$, and we arrive at the simple
relation\cite{jordan:05}
\begin{align}
   &\mathcal{R}S^{\rm irr}_{I_{\A}I_{\B}}(x_{\A},t_{\A};x_{\B},t_{\B})
   \nonumber\\
   &\approx
   \frac{E_{{\rs C}}^\A E_{{\rs C}}^\B}{\hbar^2}
   \mathcal{I}S^{\A}_{IQ,\omega=0}
   \mathcal{I}S^{\B}_{IQ,\omega=0} 
   \mathcal{R}S^{\rm irr}_{QQ}(\xi_A,\xi_B)
   \label{eq:SIIrespSQQ}
\end{align}
with $\xi_{\M}=t_{\M}-|x_{\M}|/v_{\rsF}$ for $\mathcal{M} = \mathcal{A},
\mathcal{B}$. Hence, for weak coupling and good (i.e., fast) detectors, the
measureable quantity is the symmetrized {\it irreducible} correlator
$\mathcal{R}S^{\rm irr}_{QQ} (\xi_{\A},\xi_{\B}) = \langle\langle
\{\hat{Q}(\xi_{\A}), \hat{Q}(\xi_{\B})\} \rangle\rangle/2$ which depends only
on the time-difference $\Delta \xi = \xi_{\A}-\xi_{\B}$.  Analyzing the
measurement with a single detector $\mathcal{M}$, we find that the relation
between the current-current and charge-charge correlators is of 
the form \cite{korot:01b,ave:03}
\begin{align}\label{eq:IIQQ}
   \mathcal{R}S^{\rm irr}_{I_{\M}I_{\M}}(t_{2},t_{1})
   &\approx \mathcal{R}S^{\rm irr,0}_{I_{\M}I_{\M}}(t_{2},t_{1})\\
   \nonumber
   &\quad + (E_{{\rs C}} \mathcal{I}S^{\M}_{IQ,\omega=0}/\hbar)^2 
   \mathcal{R}S^{\rm irr}_{QQ}(t_2-t_1)
\end{align}
with $\mathcal{R}S^{\rm irr,0}_{I_{\M}I_{\M}}(t_{2},t_{1})$ describing the
intrinsic current fluctuations of the detector current giving rise to a
reduced signal-to-noise ratio.
   
Finally, the measurable {\it reducible} charge-charge correlator that can be
compared to the projected charge correlator Eq.\ \eqref{eq:SPQQ} is given by
\begin{align}\label{eq:RSQQ}
   \mathcal{R}S_{QQ}(t_2-t_1)
   &=\langle\hat{Q}\rangle^2+\mathcal{R}S^{\rm irr}_{QQ}
   (t_1-t_2)\nonumber\\
   &=\langle\{\hat{Q}(t_2),\hat{Q}(t_1)\}\rangle/2.
\end{align}
As for strong coupling, spin can be taken into account by considering the full
charge operator $\hat{Q} = \hat{Q}_{\uparrow} + \hat{Q}_{\downarrow}$ in Eq.\
\eqref{eq:RSQQ} and accounting for the statistical independence of
$\hat{Q}_{\uparrow}$ and $\hat{Q}_{\downarrow}$.  

%%%%%%%%%%%%%%%%%%%%%%%%%%%%%%%%%%%%%%%%%%%%%%
\section{Charge-charge correlators}\label{sec:cc}
%%%%%%%%%%%%%%%%%%%%%%%%%%%%%%%%%%%%%%%%%%%%%%

%%%%%%%%%%%%%%%%%%%%%%%%%%%%%%%%%%%%%%%%%%%%%%
\subsection{Symmetrized charge-charge correlator}\label{sec:cc-weak}
%%%%%%%%%%%%%%%%%%%%%%%%%%%%%%%%%%%%%%%%%%%%%%

The measurable quantity at weak coupling is the symmetrized charge-charge
correlator $\mathcal{R}S_{QQ}(t_2,t_1)$. Making use of Eq.\ \eqref{eq:Qexpr}
expressing the charge $\hat{Q}(t)$ in region $\mathcal{D}$ through the matrix
${\bf A}_{k',k}(t)$ and the density matrix $\hat\rho_0$ we obtain the result
\begin{align}
\label{eq:SQQ}
   &\mathcal{R}S_{QQ}(t_2,t_1)=
   \Bigl(\sum_{\alpha}A_{\alpha,\alpha}n_{\alpha}\Bigr)^2\\
   &+\frac{1}{2}\sum_{\alpha,\beta}
   A_{\alpha,\beta}(t_2)A_{\beta,\alpha}(t_1)
   \bigl[n_{\alpha}(1-n_{\beta})
   +n_{\beta}(1-n_{\alpha})\bigr],
   \nonumber
\end{align}
with the occupation numbers $n_{\alpha}=n_{ak}=f(\epsilon_k-\mu_a)$.  Below, we will compare this correlator to the projected charge-charge correlator
$S^P_{QQ}(t_2,t_1)$.

%%%%%%%%%%%%%%%%%%%%%%%%%%%%%%%%%%%%%%%%%%%%%%%%%%%%%%%%%%
\subsection{Projected charge-charge correlator}\label{sec:cc-strong}
%%%%%%%%%%%%%%%%%%%%%%%%%%%%%%%%%%%%%%%%%%%%%%%%%%%%%%%%%%
%
Starting from the expression \eqref{eq:SPQQ} for the projected charge-charge
correlator, we make use of the Poisson formula $\sum_{N} e^{i(\lambda -
\lambda^\prime)N} = \sum_{M}2\pi \delta(\lambda-\lambda^\prime + 2\pi M)$ to
rewrite the projected charge in the form
\begin{align}
   \hat{Q}^{P}(t_2|t_1)=\int\limits_0^{2\pi}\frac{d\lambda}{2\pi}
   e^{-i\lambda\hat{Q}(t_1)}\hat{Q}(t_2)e^{i\lambda\hat{Q}(t_1)}.
\label{eq:QPt2t1}
\end{align}
Using the projector property of $\bf{A}$, see Eq.\ \eqref{eq:AAA}, we find for
the commutators (we adopt the notation in Eq.\ \eqref{eq:Qsexpr})
\begin{align}
   [e^{-i\lambda\hat{Q}(t_1)},\hat{c}_{\alpha'}^\dagger]
   &=(e^{-i\lambda}-1)\sum_{\gamma'} A_{\alpha',\gamma'}(t_1)
   \hat{c}_{\gamma'}^\dagger e^{-i\lambda\hat{Q}(t_1)},\nonumber\\
   [\hat{c}_{\alpha}^{\phantom{\dagger}},e^{i\lambda\hat{Q}(t_1)}]
   &=e^{i\lambda\hat{Q}(t_1)}(e^{i\lambda}-1)
   \sum_{\gamma} \hat{c}_{\gamma}^{\phantom{\dagger}}A_{\gamma,\alpha}(t_1),
\end{align}
and carrying out the integration in \eqref{eq:QPt2t1} over $\lambda$, the
projected charge operator can be written as
\begin{align}
   \hat{Q}^{P}(t_2|t_1)=
   \sum_{\alpha^\prime \alpha} A^{P}_{\alpha^\prime,\alpha}(t_2|t_1)
   \hat{c}^\dagger_{\alpha^\prime} \hat{c}^{\phantom\dagger}_{\alpha}.
\label{eq:QP}
\end{align}
The matrix ${\bf A}^P_{k^\prime,k}$ associated with the projected charge
assumes the form (in lead space)
\begin{align}
   &{\bf A}^P_{k^\prime ,k}(t_2|t_1)=
   \!\!\iint\!\frac{dp^\prime}{2\pi}
   \frac{dp}{2\pi} \bigl[
   {\bf A}_{k^\prime ,p^\prime}(t_1) {\bf A}_{p^\prime,p}(t_2)
   {\bf A}_{p,k}(t_1)\nonumber\\
   &\hspace{76pt}+
   {\bf \bar{A}}_{k^\prime, p^\prime}(t_1) {\bf A}_{p^\prime,p}(t_2)
   {\bf \bar{A}}_{p,k}(t_1)\bigr]
\label{eq:AP1}
\end{align}
with the matrix ${\bf A}_{k,p}(t)$ given by Eq.\ \eqref{eq:A} and ${\bf
\bar{A}}_{k,p}(t) \equiv 2\pi\delta(k-p){\bf 1} - {\bf A}_{k,p}(t)$.  Finally,
the projected reducible charge-charge correlator $S_{QQ}^P$ can be expressed
in the form
\begin{align}
\label{eq:PQQ}
   &S^P_{QQ}(t_2,t_1)
   =\sum_{\alpha}A^{P}_{\alpha,\alpha}(t_2|t_1)n_{\alpha}
    \sum_{\beta}A_{\beta,\beta}(t_1)n_{\beta} \\
   &+\frac{1}{2}\sum_{\alpha,\beta}
   A^{P}_{\alpha,\beta}(t_2|t_1)A_{\beta,\alpha}(t_1)
\bigl[n_{\alpha}(1-n_{\beta})+n_{\beta}(1-n_{\alpha})\bigr],\nonumber
\end{align}
where the first contribution corresponds to the product $\langle \hat{Q}^{P}
(t_2|t_1) \rangle \langle \hat{Q} (t_1) \rangle$.  The comparison with the
symmetrized correlator $\mathcal{R}S_{QQ}(t_2,t_1)$ found in Eq.\
\eqref{eq:SQQ} reveals quite some similarities, with one of the matrices ${\bf
A}_{k^\prime,k}$ to be replaced by its projected version ${\bf
A}^P_{k^\prime,k}$. Further below, we will analyze this correlator and compare
it to the symmetrized correlator $\mathcal{R}S_{QQ}(t_2,t_1)$.

Before doing so, we show that the projected charge matrix elements in Eq.\
\eqref{eq:AP1} can be simplified considerably.  To this end, we derive a
generalized version of Eq.\ \eqref{eq:AAA} which involves different times $t$
and $s$,
\begin{align}
   \int\frac{dp}{2\pi} {\bf A}_{k,p}(t){\bf A}_{p,q}(s)
   &=\Theta(s-t)\int\frac{dp}{2\pi} \Pi_{k,p}(t){\bf A}_{p,q}(s)
   \nonumber\\
   &\hspace{-20pt}
   +\Theta(t-s)\int\frac{dp}{2\pi} {\bf A}_{k,p}(t){\Pi}_{p,q}(s),
\label{eq:AAtoPA}
\end{align}
with the matrix elements $\Pi_{k,p}(t) = -i e^{i(k-p) (v_{\rsF} t + d/2)} /$
$(k-p-i\delta)$ and $\Theta(t)$ the Heaviside function. Combining Eqs.\
\eqref{eq:AP1} and \eqref{eq:AAtoPA} we obtain the projected charge matrix
\begin{align}
   &{\bf A}^P_{k^\prime ,k}(t_2|t_1)=
   \!\!\iint\!\frac{dp^\prime}{2\pi}
   \frac{dp}{2\pi}
   \bigl[ \Pi_{k^\prime, p^\prime}(t_1)
   {\bf A}_{p^\prime, p}(t_2)
   \Pi_{p, k}(t_1)\nonumber\\
   &\hspace{76pt}+
   \bar{\Pi}_{k^\prime, p^\prime}(t_1)
   {\bf A}_{p^\prime, p}(t_2)
   \bar{\Pi}_{p, k}(t_1) \bigr],
\label{eq:AP}
\end{align}
where $\bar{\Pi}_{k, p}(t) = 2\pi \delta(k-p)-\Pi_{k,p}(t)$.  Note the
considerable simplification of this expression as compared to the original
formula \eqref{eq:AP1}, where the structure of the region $\D$ encoded in the
matrix elements ${\bf A}_{k^\prime,k}$ through the scattering matrices ${\bf
S}_{k^\prime}^\dagger$ and ${\bf S}_{k}^{\phantom{\dagger}}$ appeared three
times, while now only one matrix ${\bf A}_{p^\prime,p}$ remains. The result
Eqs.\ \eqref{eq:QP} with \eqref{eq:AP} for the projected charge operator is
the main technical result of this paper.

The generalized projector property Eq.\ \eqref{eq:AAtoPA} is related to the
projector property of the charge operator $\hat{Q} = \hat{\mathcal{P}}_{\D}$
restricted to the single-particle Hilbert space.  Expressing the field
operators in $\int dx \, \hat{\psi}^\dagger(x)\hat{\psi}(x) = 1$ (restricted
to the single-particle Hilbert space) through the scattering states, we can
express $\hat{\mathcal{P}}_{\D}$ through the projectors onto the incoming and
outgoing parts of the scattering states,
\begin{align}\label{eq:sumP}
   \hat{\mathcal{P}}_{\D}(t) = 1-\hat{\mathcal{P}}_{\rm in}(t)
   -\hat{\mathcal{P}}_{\rm out}(t),
\end{align}
where $\hat{\mathcal{P}}_{\nu}(t)=\int dx\, \psi_{\nu}^\dagger(x,t)
\psi_{\nu}^{\phantom{\dagger}} (x,t)$ with $\nu={\rm in}/{\rm out}$. Here, we
have used that $_{\mathcal{D}}\langle\varphi_{ak} | \varphi_{bq}\rangle_{\rm
in/out}=0$ and have neglected the small overlaps $_{\rm in}\langle\varphi_{ak}
| \varphi_{bq}\rangle_{\rm out} = \mathcal{O}(1/k_{\rsF})$. Multiplying Eq.\
\eqref{eq:sumP} by $\hat{\mathcal{P}}_{\D}(s)$ with $s > t$, we obtain the
relation
\begin{align}\label{eq:PsumP}
   \hat{\mathcal{P}}_{\D}(s) \hat{\mathcal{P}}_{\D}(t)
   = \hat{\mathcal{P}}_{\D}(s) [1-\hat{\mathcal{P}}_{\rm in}(t)],
\end{align}
where we could drop the term $\hat{\mathcal{P}}_{\D}(s) \hat{\mathcal{P}}_{\rm
out}(t)$ since the outcoming component of a scattered particle cannot
contribute to the charge on the dot at a later time. Expressing
$\hat{\mathcal{P}}_{\D}$ through the charge matrix ${\bf A}_{k,p}$, see Eq.\
\eqref{eq:A}, and using  $1-\hat{\mathcal{P}}_{\rm in}(t) = \iint (dk/2\pi)
(dp/2\pi) \Pi_{k,p} (t) \hat{c}_{k}^\dagger \hat{c}_{p}$ with matrix elements
$\Pi_{k,p}$ given above, we straightforwardly arrive at Eq.\
\eqref{eq:AAtoPA}. Choosing $s < t$, we multiply Eq.\ \eqref{eq:sumP} by
$\hat{\mathcal{P}}_{\D}(s)$ from the right.

The two correlators $\mathcal{R}S_{QQ}(t_2,t_1)$ and $S^P_{QQ}(t_2,t_1)$ are
formally different, see Eqs.\ \eqref{eq:SQQ} and \eqref{eq:PQQ}. This implies
that in the two limiting measurements, a different charge dynamics is detected
due to the different back-action of the measurement onto the system. To study
the difference between these two correlators beyond a formal level, we focus
on the specific example of a single level quantum dot (modeled by the single
resonance level model\cite{oehri:12}) and investigate the differences between
$\mathcal{R}S_{QQ}(t_2,t_1)$ and $S^P_{QQ}(t_2,t_1)$ quantitatively. In the
end, we will draw some conclusions from this analysis for the general
situation.

%%%%%%%%%%%%%%%%%%%%%%%%%%%%%%%%%%%%%%%%%%%%%%
\subsection{Single resonance level model}
%%%%%%%%%%%%%%%%%%%%%%%%%%%%%%%%%%%%%%%%%%%%%%

We model a single-level quantum dot scatterer by a single resonance level with
a wavevector $k_{\rm res}$ defining its energy $\epsilon_{\rm res}=
\epsilon_{\rsF}+\hbar v_{\rsF}(k_{\rm res}-k_{\rsF})$, a width $\gamma$ (in
$k$-space), and a parameter $\eta\in[-1,1]$ describing the asymmetry in the
coupling to the two leads \cite{oehri:12}. The unprojected
charge matrix elements in Eq.\ \eqref{eq:A} assume the simple form
\begin{align}
   A_{a^\prime k^\prime,ak}(t)=
   a_{a^\prime}^\ast  a_{a}^{\phantom{\ast}} \,
   \bigl[ \phi_{k^\prime}^\ast(t)
   \phi_{k}^{\phantom{\ast}}(t) \bigr]
\label{eq:Asrl}
\end{align}
with $a_{L/R}=\pm i\sqrt{1\mp\eta}$ and $\phi_k(t)=\sqrt{\gamma}/(\delta
k+i\gamma)$ $\times e^{-i \delta k\,v_{\rsF} t}$ where $\delta k=k-k_{\rm
res}$. Alternatively, the charge operator can be expressed as
$\hat{Q}=\hat{d}^\dagger\hat{d}$ with $\hat{d}=\sum_a \int(dk/2\pi) a_a \phi_k
\hat{c}_{ak}$, where the operator $\hat{d}^\dagger$ creates a charge on the
resonance level. This model describes a single spin-degenerate dot level in
the non-interacting case with the spin trivially accounted for as noted above;
alternatively, the model applies to the case of a dot with strong Coulomb
interaction where at most one electron is present on the dot.  In the
following, we restrict ourselves to the discussion of the equilibrium case at
zero temperature, i.e., $\mu_L=\mu_R$ and $T=0$ (more precisely, we require
that $k_{\rs B} T < \max[|\epsilon_{\rm res}- \epsilon_{\rs F}|,\hbar v_{\rs
F}\gamma]$). Eq.\ \eqref{eq:Qsexpr} combined with Eq.\ \eqref{eq:Asrl} then
provide us with the equilibrium (steady-state) charge
\begin{align}
  Q&=\langle \hat{Q}(t)\rangle=
   \sum_{\alpha}A_{\alpha,\alpha}n_{\alpha}
   =\int\limits_{-\infty}^{\delta\kappa_{\rsF}}\frac{d(\delta\kappa)}{\pi}
   \frac{1}{\delta\kappa^2+1},
\label{eq:Qexps}
\end{align}
which is the measured dot occupation both for a strong- and a weak
system--detector coupling.

%
%%%%%%%%%%%%%%%%%%%%%%%%%%%%%%%%%%%%%%%%%%%%%%
\paragraph{Projected charge.}
%%%%%%%%%%%%%%%%%%%%%%%%%%%%%%%%%%%%%%%%%%%%%%
%
First, we determine the expectation value of the projected charge operator
$\hat{Q}^P(t|0)$, see Eq.\ \eqref{eq:QP}. The measurement of this quantity
involves two charge projections at times $t=0$ and later at $t>0$, where the
outcome of the first measurement is disregarded. For the single resonance
level, the projected charge matrix element Eq.~\eqref{eq:AP} is given by
\begin{align}
\label{eq:APexpl}
   &A^P_{a^\prime k^\prime,ak}(t_2|t_1)=
   a_{a^\prime}^\ast a_{a}^{\phantom{\ast}}\,
   \bigl[ \phi^{P\ast}_{k^\prime}(t_2|t_1)
   \phi^P_{k}(t_2|t_1) \\ \nonumber
   &\qquad\quad+(\phi_{k^\prime}^\ast(t_2)- \phi^{P\ast}_{k^\prime}(t_2|t_1))
   (\phi_{k}^{\phantom{\ast}}(t_2) -  \phi^P_{k}(t_2|t_1) )
   \bigr]
\end{align}
with the projected amplitude $\phi^P_{k}(t_2|t_1) = \phi_{k}^{\phantom{\ast}}
(t_1)$ $\times e^{-\gamma v_{\rsF}(t_2-t_1)}$. The expectation value of the
projected charge operator then takes the form
\begin{align}
   \langle\hat{Q}^P(t|0)\rangle \!
   =\! \int\limits_{-\infty}^{\delta\kappa_{\rsF}}\frac{d(\delta\kappa)}{\pi}
   \Bigl( \frac{e^{-2\tau}}{\delta\kappa^2+1}
   + \frac{|e^{i\delta\kappa\tau}-e^{-\tau}|^2}{\delta\kappa^2+1}\Bigr),
\end{align}
where we have introduced the dimensionless time $\tau=v_\rsF \gamma t$ and the
relative wave-vector $\delta\kappa=(k-k_{\rm res})/\gamma$. 
\begin{figure}
\begin{center}
\includegraphics[width=6.0cm]{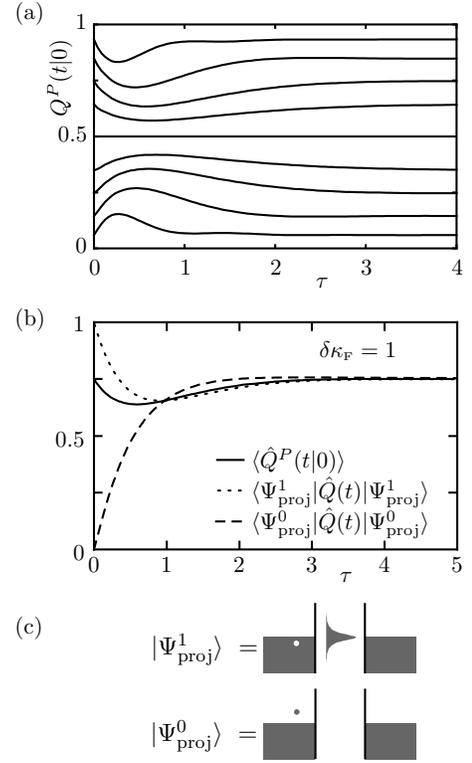}
\caption{(a) Charge expectation value $Q^{P}(t|0)$ for $\delta\kappa_{\rsF} =
\pm 5$, $\delta\kappa_{\rsF} = \pm 2$, $\delta\kappa_{\rsF} = \pm 1$,
$\delta\kappa_{\rsF} = \pm 0.5$, and $\delta\kappa_{\rsF} = 0$.  (b) Escape
dynamics of the projected electron [hole] out of the dot described by $\langle
\Psi_{\rm proj}^1| \hat{Q}(t)|\Psi_{\rm proj}^1\rangle$ (dotted) [$\langle
\Psi_{\rm proj}^0| \hat{Q}(t)|\Psi_{\rm proj}^0\rangle$ (dashed)] resulting in
the total projected charge dynamics $\langle \hat{Q}^{P}(t|0)\rangle$ (full
line) for $\delta\kappa_\rsF=1$. (c) Sketch of two possible outcomes
$|\Psi_{\rm proj}^1\rangle$ and $|\Psi_{\rm proj}^0\rangle$ of the charge
projection at $t=0$.}
\label{fig:chargeproj}
\end{center}
\end{figure}

The time dependence of $\langle\hat{Q}^P(t|0)\rangle$ for different positions
of the resonance with respect to the Fermi level is plotted in Fig.\
\ref{fig:chargeproj}. The data is (particle-hole) symmetric with respect to
the resonance's position relative to the Fermi level, with the average
projected charges below the Fermi level ($\delta\kappa_\rsF > 0$) and
symmetrically above ($-\delta\kappa_\rsF$) related via
\begin{align}
   \langle\hat{Q}^P(t|0)\rangle_{\delta\kappa_\rsF}
   =1-\langle\hat{Q}^P(t|0)\rangle_{-\delta\kappa_\rsF}.
\end{align}
This symmetry enforces the projected charge to be equal to the equilibrium
value for the resonance aligned with the Fermi level, $\langle\hat{Q}^P(t|0)
\rangle_{\delta\kappa_\rsF=0} = 1/2$ at $\delta\kappa_\rsF=0$, see Fig.\
\ref{fig:chargeproj}(a). Away from the Fermi level, the average projected
charge at time $t>0$ is suppressed in magnitude with respect to its
equilibrium value $\langle \hat{Q}^P(t=0|0) \rangle = Q$ and returns back as
$t\rightarrow \infty$ with a typical equilibration time $t_{\rm eq} \sim
1/\gamma v_\rsF |\delta\kappa_\rsF|$.

Consider, for example, a situation where the resonance is located below the
Fermi level ($\delta\kappa_\rsF>0$).  In this situation, it is more likely to
observe an electron on the dot and thus one has $Q > 1/2$ in this regime.
Surprisingly, the second projective measurement at $0 < t < t_\mathrm{eq}$
gives an average charge which is {\it smaller} than its equilibrium value (and
correspondingly for $\delta\kappa_\rsF<0$).  In order to understand the
relaxation dynamics of $\langle \hat{Q}^P(t|0) \rangle$ it is instructive to
analyze the evolution of the entire many-particle state right after the first
projection. Consider again the situation where the resonance level is placed
below the Fermi level. After the first projection, the dot is either occupied
or empty, i.e., for zero temperature the state of the system is given by
\begin{equation*}
   |\Psi_{\rm proj}^N\rangle = \frac{\hat{P}_N|\Psi_{\rm eq}\rangle}
   {\sqrt{\langle \Psi_{\rm eq}| \hat{P}_N|\Psi_{\rm eq}\rangle}}, 
   \quad N=0,~1, 
\end{equation*}
where the equilibrium state $|\Psi_{\rm eq}\rangle$ corresponds to a filled
Fermi sea, see Fig.\ \ref{fig:chargeproj}(c). Making use of the projectors
$\hat{P}_0 = \hat{d} \hat{d}^\dagger$ and $\hat{P}_1 = \hat{d}^\dagger\hat{d}$
one can see, that for both measurement outcomes $N=0,~1$ the projection
creates a single electron-hole pair in the system. The hole then screens the
excess charge on the dot and gives rise to a Friedel-type oscillation of the
charge distribution around the dot due to the sharp Fermi edge at zero
temperature.  Assume the first measurement results in an outcome $N=1$
increasing the charge value on the dot above the equilibrium value.  Since the
projection is local, the accompanying hole is also created near the dot.
During the subsequent evolution the charge in the dot can equilibrate in two
ways: 1) the electron in the dot can tunnel out above the Fermi sea, or 2) it
can tunnel out below the Fermi sea and fill the hole state. The appearance of
the second relaxation channel enhances the electron escape rate that gives
rise to the reduction of the average charge value below the equilibrium level,
see Fig.~\ref{fig:chargeproj}(b). The same picture holds for the state
$|\Psi_{\rm proj}^0\rangle$ with the excess hole in the dot compensated by an
excess electron screening the charge outside the dot. Finally, the projected
charge expectation value $\langle\hat{Q}^P(t|0)\rangle$ is a weighted average
of the two processes corresponding to the two alternatives $N=0,~1$ of the
first projection, $\langle\hat{Q}^P(t|0)\rangle = \langle \hat{Q} \rangle
\langle \Psi_{\rm proj}^1| \hat{Q}(t) |\Psi_{\rm proj}^1\rangle + (1-\langle
\hat{Q}\rangle) \langle \Psi_{\rm proj}^0| \hat{Q}(t) |\Psi_{\rm
proj}^0\rangle$, see Fig.~\ref{fig:chargeproj}(b).

The creation of the electron hole-pair due to the projection of the charge
provides energy to the system. The average energy of the excited electron-hole
pair is given by $\epsilon_{ph} =\sum_N \langle \hat{H} \hat{P}_N \rho_0
\hat{P}_N\rangle - \langle \hat{H} \rho_0 \rangle$, 
\begin{align}\label{eq:eph}
   \epsilon_{ph}
   &=2\int_{k_\rsF}^{\infty}\frac{dk}{\pi}
      \int_{-\infty}^{k_\rsF}\frac{dq}{\pi}
      \frac{(\epsilon_k-\epsilon_q)}{[\delta k^2+\gamma^2]
                                     [\delta q^2+\gamma^2]}.
\end{align}
This integral is divergent for the linear dispersion $\epsilon_k =
\epsilon_\rsF + \hbar v_\rsF(k-k_\rsF)$ [the same is true for the quadratic
dispersion as well (with $q>0$ in Eq.\ \eqref{eq:eph})], a consequence of the
description of the projection as an instantaneous process.  In an experimental
realization the measurement involves a finite time and provides a finite
amount of energy to the system, hence the particle-hole pair involves a finite
energy. E.g., for the case of single-electron projection with a strong Coulomb
coupling $E_{\rs C}>h\Gamma$ the measurement time is given by the tunneling
time $t_{\rm tun}\sim 1/\Gamma$ and the energy exchange between the system and
the detector is limited by $E_{\rs C}$; on the other hand, the measurement
process should provide the energy for a `reasonable' electron-hole pair
(defining one electron or hole in the dot) given by $\epsilon_{ph}\sim
|\epsilon_{\rsF}-\epsilon_{\rm res}|$. For a resonance overlapping with the
Fermi level $|\epsilon_{\rsF}-\epsilon_{\rm res}| < h v_{\rsF} \gamma$ and a
fast measurement with $t_{\rm tun} \ll t_{\rm sys} \sim 1/v_{\rsF} \gamma$ the
Coulomb energy exceeds the electron-hole energy as $E_{\rs C}> h\Gamma \gg h
v_{\rsF} \gamma \sim \epsilon_{ph}$. In the case of a multi-electron
projection with $E_{\rs C}\ll h\Gamma$, each individual QPC electron provides
a small amount of energy $\sim E_{\rs C}$ exciting the system to a virtual
state and the overall process creates the resulting electron-hole pair in a
similar way as the ionization via multi-photon absorption \cite{keldysh:65}.

%%%%%%%%%%%%%%%%%%%%%%%%%%%%%%%%%%%%%%%%%%%%%%
\paragraph{Charge-charge correlator.}
%%%%%%%%%%%%%%%%%%%%%%%%%%%%%%%%%%%%%%%%%%%%%%
When discussing the charge-charge correlator, it is convenient to subtract the
asymptotic value $\langle\hat{Q}\rangle^2=Q^2$ and define $\Delta
(\mathcal{R} S_{QQ}) = \mathcal{R}S_{QQ} - Q^2$ (which corresponds to
the irreducible correlator) and $\Delta S^P_{QQ} = S^P_{QQ} -Q^2$. 
For a single resonance level, the correlator $\Delta(\mathcal{R} S_{QQ})$ as
given by Eq.\ \eqref{eq:SQQ} takes the form
\begin{align}
   &\Delta(\mathcal{R}S_{QQ})(t = \tau / \gamma v_{\rsF},0)
   \label{eq:SQQsrl}
   \\ \nonumber
   &\hspace{10pt}=
   e^{-\tau}\int_{-\infty}^{\delta\kappa_{\rsF}}\!\!\frac{d(\delta\kappa)}
   {\pi} \frac{\cos(\delta\kappa\,\tau)}{\delta\kappa^2+1}
   - \Bigl| \int_{-\infty}^{\delta\kappa_{\rsF}}\!\!\frac{d(\delta\kappa)}
   {\pi} \frac{e^{i\delta\kappa\,\tau}}{\delta\kappa^2+1} \Bigr|^2.  
\end{align}
On the other hand, the projected correlator $\Delta S_{QQ}^P$ follows from
Eq.\ \eqref{eq:PQQ} and reads  
\begin{align}
   &\Delta S^P_{QQ}(t,0)=
   Q e^{-2\tau} - \Bigl|
   \int_{-\infty}^{\delta\kappa_{\rsF}}\!\!\frac{d(\delta\kappa)}{\pi}
   \frac{e^{i\delta\kappa\,\tau}}{\delta\kappa^2+1} \Bigr|^2
\label{eq:PQQsrl}
\end{align}
with the equilibrium charge expectation value $Q$ given in Eq.\
\eqref{eq:Qexps}.  Due to particle-hole symmetry, both correlators Eqs.\
\eqref{eq:SQQsrl} and \eqref{eq:PQQsrl} remain unchanged when replacing
$\delta\kappa_{\rsF} \rightarrow -\delta\kappa_{\rsF}$. Furthermore, the
difference between the two correlators is given by the first terms and
vanishes as $\delta\kappa_{\rsF} \to \pm \infty$ (filled and empty dot,
unaltered by the projection) as well as at the particle-hole symmetric point
at half-filling, $\delta \kappa_{\rsF} = 0$.  
\begin{figure}[tb]
\begin{center}
\includegraphics[width=7.5cm]{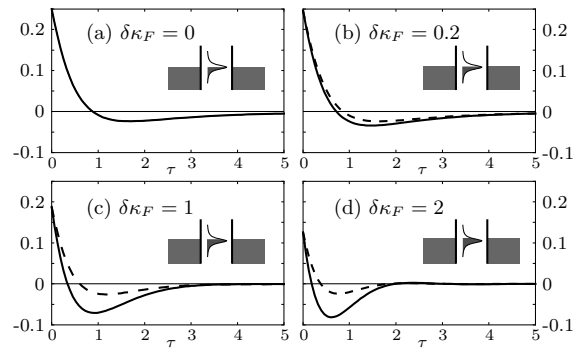}
\caption{The correlators $\Delta \mathcal{R} S_{QQ}(t,0)$
(dashed line) and $\Delta S^P_{QQ}(t,0)$ (full line) as a function of the
dimensionless time $\tau=\gamma v_{\rsF} t$ for different values of the
dimensionless resonance position $\delta\kappa_{\rsF}=(k_{\rsF}-k_{\rm
res})/\gamma$. While the two correlators are similar for small
$\delta\kappa_{\rsF}$ (see (a) $\delta\kappa_{\rsF}=0$ and (b)
$\delta\kappa_{\rsF}=0.2$), for larger $\delta\kappa_{\rsF}$, the difference
between the two correlators is more pronounced [(c) $\delta\kappa_{\rsF}=1$
and (d) $\delta\kappa_{\rsF}=2$]. The absolute value of the correlators
decreases for increasing $\delta \kappa_{\rsF}$ such that in the limit
$\delta\kappa_{\rsF}\rightarrow \infty$ both correlators vanish.}
\label{fig:QQxF0neu}
\end{center}
\end{figure}

\begin{figure}[tb]
\begin{center}
\includegraphics[width=6cm]{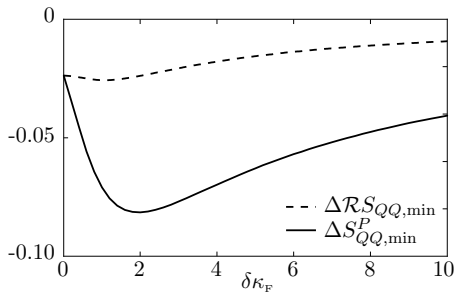}
\caption{The minima of $\Delta\mathcal{R} S_{QQ,{\rm min}} \equiv \min_t
\mathcal{R} \Delta S^P_{QQ}(t,0)$ (dashed line) and $\Delta S^P_{QQ,{\rm
min}} \equiv \min_t \Delta S_{QQ}(t,0)$ (full line) as a function of the
dimensionless resonance position $\delta\kappa_{\rsF}=(k_{\rsF}-k_{\rm
res})/\gamma$.  The anticorrelation dip is much more pronounced for the
projected correlator.}
\label{fig:SQQmin}
\end{center}
\end{figure}

Analyzing the time dependence of the two correlators, see Figs.\
\ref{fig:QQxF0neu} (a)-(d), we observe that both correlators become negative
at $t \sim 1/\gamma v_{\rsF}|\delta\kappa_\rsF|$ (cut off at $t \sim 1/\gamma
v_\rsF$ as $|\delta\kappa_\rsF|$ drops below unity) indicating the
anti-correlation in the system: detecting a particular charge value on the dot
at $t=0$, the observed charge state of the dot after the tunneling time is
more likely to be inverse. At large times both correlators show a weak
oscillating behavior approaching $\Delta \mathcal{R} S_{QQ} (t\to\infty,0) =
\Delta S^P_{QQ} (t\to\infty,0)\rightarrow 0$ and the full charge-charge
correlator assumes its asymptotic value $Q^2$. Comparing the two correlators
with one another, we identify two regimes: (1) For $\delta\kappa_{\rsF}
\lesssim 1$, the system is close to the particle-hole symmetric point and the
time-dependence of the two correlators is qualitatively the same with only a
small quantitative difference. (2) When the Fermi level is away from the
resonance level, the two correlators deviate considerably as the projection
enhances the non-equilibrium behavior of the charge at later times. Indeed,
the projected correlator shows a larger anti-correlation and more pronounced
oscillations than the symmetric one, see Fig.\ \ref{fig:SQQmin}.

The following general conclusions can be drawn from the above results: If the
region $\mathcal{D}$ is in a well-defined charge state, the two correlators
are identical as the projection due to the first charge measurement does not
alter the state. For small deviations from a well-defined charge state, the
difference between the two correlators remains small. The same is true for a
system with particle-hole symmetry, as discussed for the half-filled dot
above. If the system is in a superposition of different charge states with
similar weights (but away from a particle-hole symmetric point), the
projection has a pronounced impact on the measurement outcome and the two
correlators $\mathcal{R} S_{QQ}(t,0)$ and $S^P_{QQ}(t,0)$ can deviate
considerably with specific details depending on the scattering matrix of the
system.

%%%%%%%%%%%%%%%%%%%%%%%%%%%%%%%%%%%%%%%%%%%%%%
\section{Effect of multiple projective measurements}\label{sec:mp}
%%%%%%%%%%%%%%%%%%%%%%%%%%%%%%%%%%%%%%%%%%%%%%

So far, we have discussed two regimes of measuring the time-resolved
charge-charge correlator: i) by projections (strong measurements) at times
$t_1$ and $t_2$, or ii) by a weak measurement where the charge-charge
correlator is obtained from a deconvolution of the detector current-current
correlator and the detector response function.  While in the latter case, the
detector current may be applied continuously, for the strong measurement it is
crucial to measure the charge at two times in a strong manner and allow for an
unperturbed evolution in-between.  Let us compare these two regimes to a
typical experimental situation \cite{gust:06}: In these experiments, a fixed
voltage $V$ is applied at the detector which thus is constantly monitoring the
charge of the system; at the same time, the clear steps visible in the
detector-current traces indicate a strong measurement. Given the constant
monitoring of the dot charge, the system is always perturbed.  Furthermore,
while our analysis above assumes a fully coherent evolution of the system, the
coupling to the environment leads to a further decoherence beyond the one
introduced by the measurement.  Our model of a projective measurement of the
charge-charge correlator $S^P_{QQ} (t_2,t_1)$ with an initial projection at
time $t_1$ and a final projection at time $t_2$ is inappropriate to describe
these measurements.

In the following, we formulate the continuous strong measurement discussed
above and derive the appropriate charge-charge correlator. Following the
general spirit of the paper, we model the continuous strong measurement by
repeated projective measurements of the system. We introduce the formalism of
multiple projective measurements and present the results for the multiple
projected charge expectation value as well as the multiple projected
charge-charge correlator. We identify a critical measurement rate where the
dynamics of the charge changes from a system property to a regime where it is
dominated by the measurement, becoming universal in the limit of a high
projection rate. In the end, we discuss the rate at which we should project
the system in order to model a realistic continuous measurement and compare to
experiments.

%%%%%%%%%%%%%%%%%%%%%%%%%%%%%%%%%%%%%%%%%%%%%%
\subsection{Charge expectation value}\label{sec:multprojQ}
%%%%%%%%%%%%%%%%%%%%%%%%%%%%%%%%%%%%%%%%%%%%%%

Starting from the situation where we project the density matrix $\hat{\rho}_0$
once, see Sec.\ \ref{sec:cc-strong} above, we account for $n$ projections at
times $t_1 < \ldots < t_n$ via the multiply projected density matrix
\begin{align}
   \sum_{N_1,\ldots,N_n}
   \mathcal{T}\Bigl[\prod_{i=1}^{n}\hat{P}_{N_i}(t_i)\Bigr]
   \hat{\rho}_0
   \tilde{\mathcal{T}}\Bigl[\prod_{i=1}^{n}\hat{P}_{N_i}(t_i)\Bigr]
\end{align}
with the time-ordering operator $\mathcal{T}$ and the anti-time-ordering
operator $\tilde{\mathcal{T}}$.  The charge expectation at time $t>t_n$ after
$n$ previous projections at times $t_1 < \ldots < t_n$ is given by the
expression
\begin{align}
   &\sum_{N_1,\ldots,N_n} \text{Tr}\Bigl\{
   \hat{Q}(t)\mathcal{T}\Bigl[\prod_{i=1}^{n}\hat{P}_{N_i}(t_i)\Bigr]
   \hat{\rho}_0
   \tilde{\mathcal{T}}\Bigl[\prod_{i=1}^{n}\hat{P}_{N_i}(t_i)\Bigr]
   \Bigr\}\nonumber\\
   &=\sum_{N_1,\ldots,N_n} \text{Tr}\Bigl\{
   \tilde{\mathcal{T}}
   \Bigl
   [\prod_{i=1}^{n}\hat{P}_{N_i}(t_i) \Bigr] \hat{Q}(t)\mathcal{T}
   \Bigl [\prod_{i=1}^{n}\hat{P}_{N_i}(t_i) \Bigr] \hat{\rho}_0
   \Bigr\} \nonumber\nonumber\\
   &
   \equiv\langle\hat{Q}^{mP}(t|\{t_j\})\rangle,
\end{align}
where we have used again the trace property ${\rm Tr}(\hat{A}\hat{B})={\rm Tr}
(\hat{B}\hat{A})$ and we have introduced the multiply projected charge
operator $\hat{Q}^{mP}(t|\{t_j\})$ with the projection times $\{t_{j}\} \equiv
t_1, \ldots, t_n$. The latter can be expressed as $\hat{Q}^{mP} (t|\{t_j\}) =
\sum_{\alpha^\prime, \alpha} A^{mP}_{\alpha^\prime, \alpha} (t|\{t_j\})
\hat{c}^\dagger_{\alpha^\prime}\hat{c}^{\phantom{\dagger}}_{\alpha}$ with
matrix elements $A^{mP}_{\alpha^\prime, \alpha}(t|\{t_j\})$. To obtain these
matrix elements, the projection operators can be taken into account
iteratively, i.e.,  $\hat{Q}^{mP} (t|t_n, \ldots, t_i) = \sum_{N_i}
\hat{P}_{N_i} (t_i) \hat{Q}^{mP} (t|t_n, \ldots, t_{i+1}) \hat{P}_{N_i}(t_i)$.
Making use of Eq.~\eqref{eq:AAtoPA} and the property $\int (dq/2\pi) \Pi_{pq}
(t) \Pi_{qk}(t') = \Pi_{pk} (t_{\rm min})$ with $t_{\rm min}=\min(t,t')$, the
resulting matrix element can be simplified to
\begin{align}
   &\hat{\bf A}^{mP}_{k^\prime ,k}(t|\{t_j\})
   =\iint\!\frac{dp^\prime}{2\pi}\frac{dp}{2\pi}
   \Bigl[
   \Pi_{k^\prime p^\prime}(t_1)
   {\bf A}_{p^\prime,p}(t)
   \Pi_{pk}(t_1)
   \nonumber\\
   &\quad+\sum_{i=1}^{n-1}\bigl[\Pi(t_{i+1})-\Pi(t_{i})\bigr]_{k^\prime
   p^\prime}
   {\bf A}_{p^\prime,p}(t)
   \bigl[\Pi(t_{i+1})-\Pi(t_{i})\bigr]_{pk}\nonumber\\
   &\quad+
   \bar{\Pi}_{k^\prime p^\prime}(t_n)
   {\bf A}_{p^\prime,p}(t)
   \bar{\Pi}_{pk}(t_n)
   \Bigr].
\label{eq:Amultpr}
\end{align}
%

%%%%%%%%%%%%%%%%%%%%%%%%%%%%%%%%%%%%%%%%%%%%%%
\subsubsection{Single resonance level model}
%%%%%%%%%%%%%%%%%%%%%%%%%%%%%%%%%%%%%%%%%%%%%%

\begin{figure}
\begin{center}
\includegraphics[width=7cm]{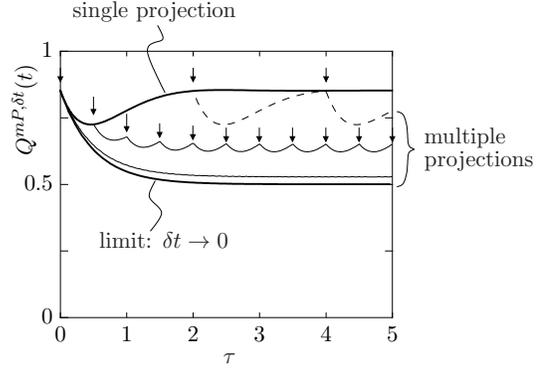}
\caption{For a single initial projection, the charge $Q^{P}(t|0)$ returns to
the equilibrium value after a characteristic time scale $t_{\rm eq}$ (upper
thick line, we choose $\delta\kappa_{\rs F}=(k_{\rs F}-k_{\rm
res})/\gamma=2$).  For multiple projections (indicated by arrows) with time
steps $\delta t$, we find two different behaviors for the charge $Q^{mP,\delta
t}(t)$: For $\delta t>t_{\rm eq}$, the dot returns to the equilibrium before
the next projection (dashed line, $\delta t= 2/\gamma v_{\rs F}$), while this
is not the case when $\delta t<t_{\rm eq}$ and the system is driven to a new
steady state (upper thin line, $\delta t= 1/2\gamma v_{\rs F}$). For faster
projections (lower thin line, $\delta t= 1/10\gamma v_{\rs F}$), the
occupation value is driven towards 1/2 (with the limit $\delta t\rightarrow 0$
indicated by the lower thick line). }
\label{fig:chargemultproj}
\end{center}
\end{figure}

The relevant timescale in the description of (multiple) projections of the
mean charge is the equilibration time $\tau_{\rm eq}=\gamma v_\rsF t_{\rm eq}
\sim 1/\delta\kappa_\rsF$ (for $\delta\kappa_\rsF\gg 1$).  For multiple
projections, there are two regimes: if the time between subsequent projections
$\delta t = \delta \tau/ v_{\rs F}\gamma$ exceeds the equilibration time,
$\delta t > t_{\rm eq}$, the projections are independent, see Fig.\
\ref{fig:chargemultproj} (dashed line) for a plot of the corresponding
multiply projected charge $Q^{mP,\delta t}(t) \equiv \langle \hat{Q}^{mP}
(t|\{t_j = (j-1)\delta t\})\rangle$. On the other hand, for short intervalls
$\delta t < t_{\rm eq}$, the system is not in equilibrium when the next
projection occurs; multiple charge projections with small equal time steps
$\delta t$ then drive the system to a new steady state, see Fig.\
\ref{fig:chargemultproj} (thin line).

When the Fermi energy resides above the resonance level ($\delta \kappa_{\rsF}
> 0$), the new 'steady state' corresponds to a lower occupation of the dot
state than in equilibrium, while for a resonance level above the Fermi energy
($\delta\kappa_{\rsF}<0$), the dot occupation is increased. Frequent
projection pushes the steady state occupation towards 1/2, see Fig.\
\ref{fig:chargemultproj}.  For $|\delta\kappa_{\rsF}| \delta \tau\ll \pi$, the
steady state occupation (at the times of projection) is given by
\begin{align}
   \lim_{\delta t\rightarrow 0}Q^{mP,\delta t}(t)
   =\frac{1}{2}+\frac{\delta\kappa_{\rsF}\delta\tau}{2\pi},
\label{eq:onehalf}
\end{align}
i.e., the dot occupation is 1/2 up to a small correction. We can attribute
this universal  behavior (independent of applied chemical potentials) to an
effective `broadening' of the level over the width $h/\delta t$ in energy---as
this width becomes much larger than the distance $\hbar v_\rsF |\delta
k_\rsF|$ of the resonance level to the Fermi level, the dot is filled and
emptied with equal probability. 

%%%%%%%%%%%%%%%%%%%%%%%%%%%%%%%%%%%%%%%%%%%%%%
\subsubsection{Comparison to experiment}
%%%%%%%%%%%%%%%%%%%%%%%%%%%%%%%%%%%%%%%%%%%%%%

Let us compare these results and in particular the appearance of a universal
regime with a charge expectation value 1/2 to the experiments by Gustavsson
{\it et al.} \cite{gust:06}; as it turns out, this experiment is not dominated
by the measurement but rather by temperature and no universal behavior is
expected. The experimental setup consists of a QPC detector which is measuring
the charge states $|N\rangle$ or $|N+1\rangle$ of a capacitively coupled
quantum dot. The QPC detector is driven with a constant voltage $V_{\rs QPC}
\sim 500{\,\rm \mu V}$ producing a current $I_{\rm in} \sim 20 {\,\rm nA}$
which corresponds to electrons impinging on the QPC at a rate $1/t_{\rm V}\sim
10^{11} {\,\rm Hz}$.  From the current steps, we conclude that the
transmissions corresponding to the two charge states are given by
$T_{|N\rangle} \sim 0.3$ and $T_{|N+1\rangle} \sim 0.2$, implying a coupling
$E_{\rs C}/h\Gamma \sim 0.1$.  Making use of Eq.\ \eqref{eq:tms}, we obtain
that $N=t_{\rm ms}/t_V \sim 20$ electrons determine the dot state.  For each
data point, the QPC current is integrated during $t_{\rm dp} \sim 50 {\,\rm
\mu s}$ such that $6.  \, 10^6$ electrons pass through the QPC, which is far
larger than $N_{\rm proj}\sim 20$. With $t_{\rm dp} < t_{\rm sys}$ every data
point then provides the information of the actual charge state of the dot. The
system lifetime derives from the step widths (of order $t_{\rm sys} \sim 1$
ms) of the measurement tracks and amounts to $h/t_{\rm sys} \sim 10^{-5}
{\,\rm \mu eV}$.  We model the effect of the continuous measurement by
repeated projections on the time scale $\delta t \sim t_{\rm ms}$, that
provides an effective level broadening $h/t_{\rm ms} \sim 2 {\,\rm \mu eV}$.
This value is far below the temperature scale \cite{gust:06} $k_{\rs B} T \sim
20 {\,\rm \mu eV}$ (at $T \sim 230 {\,\rm mK}$). Alternatively, the system is
drive-dominated when $V_{\rs QD} > k_{\rs B} T$, see Ref.\
\onlinecite{gust:06}.

%%%%%%%%%%%%%%%%%%%%%%%%%%%%%%%%%%%%%%%%%%%%%%
\subsection{Charge-charge correlator}
%%%%%%%%%%%%%%%%%%%%%%%%%%%%%%%%%%%%%%%%%%%%%%

Next, we investigate the effect of multiple projections on the charge-charge
correlator. We consider the correlator between charges at times $t_1$ and $t$
with multiple projections at times $t_i$ in-between, assuming a free evolution
of the system at times prior to the first measurement at $t_1$.  The
corresponding correlator is given by
\begin{align}
   &S_{QQ}^{mP}(t,t_1|\{t_j\})
   \nonumber\\
   &
   =\!\!\!
   \sum_{N_1,\ldots,N_n}\!\!\!  \text{Tr}\Bigl\{
   \hat{Q}(t)\mathcal{T}\Bigl[\prod_{i=1}^{n}\hat{P}_{N_i}(t_i)\Bigr]
   \hat{Q}(t_1)\hat{\rho}_0
   \tilde{\mathcal{T}}\Bigl[\prod_{i=1}^{n}\hat{P}_{N_i}(t_i)\Bigr] \Bigr\}
   \nonumber\\&
   =\text{Tr}\Bigl\{ \hat{Q}^{mP}(t|\{t_j\}) \hat{Q}(t_1)\hat{\rho}_0
   \Bigr\},
\end{align}
where we have used the trace property and the definition of the multiply
projected charge operator $\hat{Q}^{mP}(t|\{t_j\})$. Making use of the matrix
elements $A^{mP}_{\alpha,\beta}(t|\{t_j\})$ in Eq.\ \eqref{eq:Amultpr}, the
charge-charge correlator can be expressed as
\begin{align}\label{eq:Smultproj}
   &S_{QQ}^{mP}(t,t_1|\{t_j\})
   = \! \sum_{\alpha} \! A^{mP}_{\alpha,\alpha}(t|\{t_j\})n_{\alpha}
   \! \sum_{\beta} \! A_{\beta,\beta}(t_1)n_{\beta}  \\
   &+\frac{1}{2}
   \sum_{\alpha,\beta}
   \! A^{mP}_{\alpha,\beta}(t|\{t_j\})
   A_{\beta,\alpha}(t_1)
   \bigl[n_{\alpha}(1\! -\! n_{\beta})+n_{\beta}(1\! -\! n_{\alpha})\bigr],
   \nonumber
\end{align}
where the first contribution is $\langle\hat{Q}^{mP}(t|\{t_j\})\rangle
\langle\hat{Q}(t_1)\rangle$.  Comparing this result to the case of a single
projection discussed before, see Eq.~\eqref{eq:PQQ}, we re\-cognize that the
multiple projections enter through the matrix elements
$A^{mP}_{\alpha,\beta}(t|\{t_j\})$ replacing the (singly projected) matrix
element $A^{P}_{\alpha,\beta}(t|0)$. 

%%%%%%%%%%%%%%%%%%%%%%%%%%%%%%%%%%%%%%%%%%%%%%
\subsubsection{Single resonance level model}
%%%%%%%%%%%%%%%%%%%%%%%%%%%%%%%%%%%%%%%%%%%%%%

\begin{figure}
\begin{center}
\includegraphics[width=7cm]{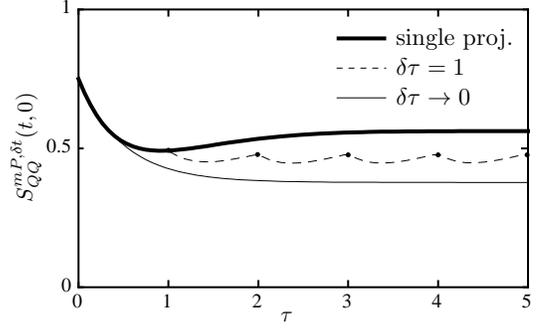}
\caption{$S^{mP,\delta t}_{QQ}(t,0)$ for multiple
projections at $\delta\kappa_{\rsF} = 1$: Single projected correlator (thick
line), multiply projected correlator (dashed line, with $\delta\tau=1$ and
dots indicating the projections), and the limiting behavior for $\delta
t\rightarrow 0$ (thin line).}
\label{fig:PPQmultproj}
\end{center}
\end{figure}

For the single resonance level in equilibrium ($\mu_L=\mu_R=\epsilon_{\rsF}$)
the charge matrix element assumes the simple form in Eq.\ \eqref{eq:Asrl}. In
Fig. \ref{fig:PPQmultproj} we compare the projected correlator $S_{QQ}^{P}
(t,0)$ with the multiply projected correlator $S_{QQ}^{mP,\delta t} (t,0)
\equiv S_{QQ}^{mP} (t,t_1=0 | \{t_j=(j-1) \delta t\})$ with equal time
separations $\delta t = \delta \tau /\gamma v_{\rsF}$ between projections; we
consider the cases $\delta \tau = 1$ and $\delta \tau \rightarrow 0$.  All
correlators start out at $t = 0$ with the value $Q$ of the average charge Eq.\
\eqref{eq:Qexps}. The correlator $S_{QQ}^{P}(t,0)$ then
undergoes a pronounced anti-correlation dip and approaches $\langle \hat{Q}(t)
\rangle \langle \hat{Q}(0) \rangle = Q^2$ for $t\rightarrow \infty$ (as the
two charge states at $0$ and $t$ become uncorrelated). Repeated projections in
between drive the correlator $S_{QQ}^{mP,\delta t}(t,0)$ towards $\langle
\hat{Q}^{mP,\delta t} (t)\rangle \langle \hat{Q}(0) \rangle$ as the
measurement induces an additional time dependence.  In the limit of frequent
projections, the projected charge expectation value $\langle\hat{Q}^{mP,\delta
t} (t)\rangle$ approaches 1/2 and hence the asymptotic value of the correlator
is given by $S_{QQ}^{mP,\delta t\rightarrow 0}(t,0) \sim Q/2$.  Overall, the
result shows that only small times $\tau < 1$ probe the system dynamics, while
larger times become dominated by the measurement.

%

%%%%%%%%%%%%%%%%%%%%%%%%%%%%%%%%%%%%%%%%%%%%%%
\subsection{Fluctuating quantum dot level}\label{sec:osclev}
%%%%%%%%%%%%%%%%%%%%%%%%%%%%%%%%%%%%%%%%%%%%%%

\begin{figure}
\begin{center}
\includegraphics[width=7cm]{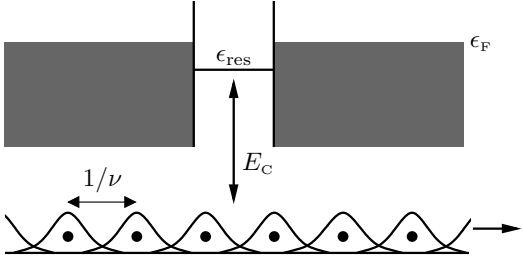}
\caption{A single resonance level at energy $\epsilon_{\rm res}$ coupled to
two leads with states filled up to the Fermi energy $\epsilon_{\rsF}$ is
capacitively interacting (with interaction strength $E_{\rs C}$) with the
electrons passing through the QPC with frequency $\nu$. We take this
interaction into account via an effective time-dependent level energy
$\epsilon_{\rm res}(t)$.}
\label{fig:osc_level}
\end{center}
\end{figure}

We have seen that repeated projections with a time separation between
subsequent projections approaching the Heisenberg time $h / |\epsilon_{\rs F}
- \epsilon_{\rm res}|$ start dominating the dynamic evolution of the system
and the average occupation number of the dot approaches the universal value
$\langle \hat{Q} \rangle = 1/2$, irrespective of the position of the dot level
with respect to the Fermi level, see Eq.\ \eqref{eq:onehalf}. Here, we show
how this formal result of the von Neumann projection postulate can be
understood in terms of a measurement-induced fluctuation of the dot level. As
before, we assume that the state of the dot is measured by a capacitively
coupled QPC, see Fig.\ \ref{fig:osc_level}. We assume single electrons passing
through the QPC with a frequency $\nu$, e.g., due to an applied voltage $V$
corresponding to $\nu=eV/h$, which corresponds to an incoming QPC current
$I=e\nu$.  An electron passing through the QPC interacts typically during a
time $t_{\rm tun}=1/\Gamma$ with the dot. The interaction strength is given by
the Coulomb energy $E_{\rs C}$.  In our model, we take into account the effect
of this current on the dot level $\epsilon_{\rm res}$ via a classical
electrostatic potential of the form $\epsilon_{\rm res}(t) = \epsilon_{\rm
res} + \tilde{E}_{\rs C}\cos(2\pi\nu t)$ with an effective coupling strength
$\tilde{E}_{\rs C}=E_{\rs C} \nu/\Gamma$ accounting for the typical
interaction time $t_{\rm tun}=1/\Gamma$. In order to describe the system, we
define the time-dependent Hamiltonian
\begin{align}
   \hat{H}(t) &=  \epsilon_{\rm res}(t) \hat{c}^\dagger\hat{c}
   +\int\frac{dk}{2\pi} \epsilon_k (\hat{a}_k^\dagger 
    \hat{a}^{\phantom{\dagger}}_k
   + \hat{b}_k^\dagger \hat{b}^{\phantom{\dagger}}_k)
   \\ & \!\!+ \int\!\!\frac{dk}{2\pi} \bigl[\hbar v_{\rsF} \sqrt{\gamma_L} 
   (\hat{a}_k^\dagger \hat{c} + h.c.)
   + \hbar v_{\rsF}\sqrt{\gamma_R} (\hat{b}_k^\dagger \hat{c} + h.c.)\bigr],
   \nonumber
\end{align}
where $\hat{a}_k$, $\hat{b}_k$ are the annihilation operators for the left and
right reservoirs, $\hat{c}$ is the annihilation operator corresponding to the
localized dot state, $\gamma_{L,R}$ are the partial widths of the level,
$\epsilon_k = \epsilon_{\rsF} + \hbar v_{\rsF} (k-k_{\rsF})$ is the linearized
dispersion of the lead electrons, and $\epsilon_{\rm res}(t)$ is the
fluctuating energy of the dot.  Solving the Heisenberg equation of motion, we
obtain
\begin{align}
   \hat{c}(t) &= \hat{c}(t_0)e^{-i\phi(t) 
   + i\phi(t_0) -\gamma v_\rsF(t-t_0)/2}\\
   &\quad -i  v_{\rsF} \int\frac{dk}{2\pi} 
   (\sqrt{\gamma_L} \hat{a}_k + \sqrt{\gamma_R}\hat{b}_k)
   \nonumber\\
   &\hspace{20pt}
   \times\int\limits_{t_0}^t dt^\prime 
   e^{-i\phi(t) + i\phi(t^\prime)- \gamma v_\rsF(t-t^\prime)/2 
   -i\epsilon_k(t^\prime - t_0)/\hbar},
   \nonumber
\end{align}
where $\gamma = (\gamma_L+\gamma_R)$ is the resonance width and $\phi(t) =
\int^t dt^\prime \epsilon_{\rm res}(t^\prime)/\hbar$. Taking $t_0 \rightarrow
-\infty$ and $t = 0$, the steady-state dot occupation is given by
\begin{align}
   \langle \hat{Q}\rangle &= \langle \hat{c}^\dagger(0)
   \hat{c}(0) \rangle = v_{\rsF}^2 \int\frac{dk}{2\pi} 
   (\gamma_L n_{Lk} + \gamma_R n_{Rk})
   \\
   &\hspace{5pt}\times\int\limits_{-\infty}^0 dt_1 dt_2
   e^{-i\phi(t_1) +i\phi(t_2) + \gamma v_\rsF (t_1+t_2)/2 
   + i\epsilon_k(t_1-t_2)/\hbar},
   \nonumber
\end{align}
which in equilibrium, i.e., $n_{Lk} = n_{Rk} = \Theta(k_{\rs F}-k)$,
simplifies to
\begin{equation*}
   \langle \hat{Q}\rangle = \gamma v_{\rsF} \!\!\!\!
   \int\limits_{-\infty}^0 \!\!\!  \frac{dt_1
   dt_2}{2\pi i} \frac{e^{-i\phi(t_1) + i\phi(t_2)}}{t_1-t_2-i\delta}\,
   \! e^{\gamma v_{\rsF} (t_1+t_2)/2}e^{i\epsilon_{\rsF}(t_1-t_2)/\hbar}\!.
\end{equation*}
The phase $\phi(t)$ is of the form $\phi(t)=\epsilon_{\rm res} \, t/\hbar +
(\tilde{E}_{\rs C}/h\nu) \sin(2\pi \nu t)$ and using the expansion
$e^{iz\sin\theta} = \sum_n J_n(z)e^{in\theta}$, we can rewrite the expression
above as
\begin{align}
   &   \langle \hat{Q}\rangle = \sum_{n,m=-\infty}^{\infty} \!\!\!\!
   J_n\Bigl( \frac{\tilde{E}_{\rs C}}{h\nu}\Bigr)
   J_m\Bigl( \frac{\tilde{E}_{\rs C}}{h\nu}\Bigr)
   \gamma v_\rsF \!\! \int\limits_{-\infty}^0\!\! dt\,
   e^{\gamma v_\rsF t -i2\pi(n-m)\nu t}
   \nonumber\\
   &\qquad\qquad\times
   \int\limits^{-2t}_{2t} \frac{d\tau}{2\pi i}\,
   \frac{e^{-i2\pi(n+m)\nu\tau +i(\epsilon_{\rs F}
   -\epsilon_{\rm res})\tau/\hbar}}{\tau-i\delta}
\end{align}
with $J_n(z)$ the Bessel functions of the first kind. Assuming frequent
projections, $h\nu \gg |\epsilon_{\rs F}-\epsilon_{\rm res}|$, and a narrow
resonance, $\hbar\gamma v_{\rsF} \ll |\epsilon_{\rs F}-\epsilon_{\rm res}|$,
we can calculate $\langle \hat{Q}\rangle$ to leading order, where only terms
$n=m\leq 0$ contribute, and obtain
\begin{equation}
   \langle \hat{Q}\rangle = \frac12\Bigl[ 1 + \text{sgn}(\epsilon_{\rs F}
   -\epsilon_{\rm res})\, J_0^2
   \Bigl(\frac{\tilde{E}_{\rs C}}{h\nu}\Bigr)\Bigr],
\end{equation}
where the identity $1 = [J_0(x)]^2+2\sum^{\infty}_{\nu=1}[J_{\nu}(x)]^2$ has
been used. Making use of $\tilde{E}_{\rs C}/h\nu=E_{\rs C}/h\Gamma$, we notice
that for $E_{\rs C}>h\Gamma$, the steady-state charge $\langle \hat{Q}\rangle$
approaches the asymptotic value $1/2$.  Hence, assuming a strong
system-detector coupling with each electron realizing a strong projective
measurement, we reproduce the case of frequent projective measurements
discussed before, see Eq.\ \eqref{eq:onehalf}. In order to reproduce the case
of a projection by $N$ electrons within a time $t_{\rm ms}$, we consider a
periodic oscillation $\epsilon_{\rm res}(t) = \epsilon_{\rm res} +
\tilde{E}_{\rs C} \cos(2\pi \tilde{\nu} t)$ of the dot level with frequency
$\tilde{\nu} = \nu/N = 1/t_{\rm ms}$ and effective Coulomb strength
$\tilde{E}_{\rs C} = N E_{\rs C}\tilde{\nu}/\Gamma = E_{\rs C}\nu/\Gamma$.
While this periodicity is artificially introduced here, it could be realized
by switching the QPC current on and off repeatedly with this period. Adapting
the calculation above, we obtain in the limit of fast projection $h\tilde{\nu}
\gg |\epsilon_{\rs F}-\epsilon_{\rm res}|$ an average charge $\langle
\hat{Q}\rangle = (1/2)[ 1 + \text{sgn}(\epsilon_{\rs F} -\epsilon_{\rm res})\,
J_0^2 (NE_{\rs C}/h\Gamma)]$, such that for $NE_{\rs C}>h\Gamma$ the system is
driven to the universal regime as in the case of single electron projection
discussed above.  Within this treatment, it is the randomization of the dot
level which leads to equal probabilities to find the level either empty or
filled and which is responsible for the universal filling factor 1/2.

%%%%%%%%%%%%%%%%%%%%%%%%%%%%%%%%%%%%%%%%%%%%%%
\section{Conclusion}\label{sec:con}
%%%%%%%%%%%%%%%%%%%%%%%%%%%%%%%%%%%%%%%%%%%%%%

We have studied the charge and charge-charge correlator of a quantum dot
$\mathcal{D}$ (a localized region $\mathcal{D}$ of an arbitrary mesoscopic
scatterer in more general terms).  Making use of the single-particle
scattering matrix of the scatterer ${\bf S}_{k}$, we expressed the charge
operator $\hat{Q}(t)$ for $\mathcal{D}$ through the single-particle scattering
matrix. In order to measure this charge, the system has been coupled to a
detector; the measurement then acts back on the dot's occupation and therefore
the charge dynamics depends on the coupling to the detector.  We have focussed
on a QPC detector which is capacitively coupled to the charge in region
$\mathcal{D}$.  We have analyzed different measurement regimes and identified
two particular cases: at strong dot--detector coupling $E_{\rs C} > h\Gamma$,
one single electron already can perform a strong measurement and project the
system; such a strong/projective measurement is still possible even at
intermediate coupling $E_{\rs C} < h\Gamma$ where many electrons accumulated
over a measurement time $t_\mathrm{ms}$ are needed to project the dot-state,
provided that $t_{\rm ms}$ remains below the system time $t_{\rm sys}$ where
the charge on the dot changes.  In our phenomenological approach, we have
described such strong measurements with the help of the von Neumann projection
postulate (maximal back action) and the measured charge-charge correlator is
the projected quantity $S^{P}_{QQ} (t,0)$ involving the projected charge
operator $\hat{Q}^P(t|0)$.  When the coupling is weak, $E_{\rs C} \ll h\Gamma$
and $t_{\rm ms}>t_{\rm sys}$ the measurement is in the weak regime.  We have
analyzed this case in lowest-order perturbation theory (with vanishing back
action) and have identified the symmetrized irreducible correlator
$\mathcal{R} S_{QQ} (t,0)$ as the measurable quantity.

The difference between the strong and weak measurement regimes consists in the
different degree the system becomes entangled with the detector during the
systems's dynamical time.  In a strong measurement, the projection of the
detector degree of freedom implies a projection of the system's state,
allowing to exclude the detector from the consideration. For the weak
measurement, the system becomes only weakly entangled with the detector and
almost preserves its coherent evolution; the detector's projection (involving
a quasi-classical variable \cite{lesovik:98}) then acts only as a weak
perturbation on the system's dynamics.  Although the results for strong and
weak measurements look quite different on a formal level, we find a
qualitative correspondence when analyzing the expressions for a single-level
quantum dot.

The projected charge correlator involves the projected charge operator
$\hat{Q}^P(t|0)$ providing the charge at time $t$ after a projection at an
earlier time $t=0$, see Eq.\ \eqref{eq:QP}.  A simple expression for this
quantity (that could also be carried over to the case of multiply repeated
projections) was obtained by making use of the analytic structure of the
scattering matrix ${\bf S}_k$. The evolution of this projected charge, in
particular, its initial decay in modulus and subsequent return to the
equilibrium value, could be understood in terms of the generation of a
particle-hole excitation in the nearby leads through the projection. In the
limit of frequent projections, we have identified a transition from a regime
where the dot dynamics is determined by the characteristics of the system to a
regime where it is uniquely determined by the detector with universal outcome.
In order to reproduce these results within a unitary treatment, we have
modeled the projective measurements by a strongly fluctuating dot level.

The back-action of a measurement naturally manifests itself in the
time-correlator of the measured quantity, as the second measurement probes
both the dynamics of the system as well as the back-action on the system
originating from the first measurement. Furthermore, the strength of the
back-action increases from zero in a weak measurement to a maximum in a strong
measurement described by the von Neumann projection scheme.  Our analysis of
time-correlators then provides insights on the range of back-action effects in
a simple model setup.

We thank Simon Gustavsson and Oded Zilberberg for illuminating discussions and
acknowledge financial support from the Swiss National Science Foundation
through the National Center of Competence in Research on Quantum Science and
Technology (QSIT), the Pauli Center for Theoretical Studies at ETH Zurich, and
the RFBR Grant No.\ 14-02-01287.

\end{document}